\documentclass[prb,reprint,aps,amsmath,amssymb,twocolumn,superscriptaddress]{revtex4-2}
\usepackage{graphicx}
\usepackage{dcolumn}
\usepackage{amsmath,bm}
\usepackage{amssymb}
\usepackage[dvipsnames]{xcolor}
\usepackage{float}
\usepackage[colorlinks=true,allcolors=Blue,linktocpage=true]{hyperref}
\usepackage{makecell}
\usepackage{calrsfs}
\usepackage{enumitem}
\usepackage{placeins}
\usepackage{soul}
\usepackage{CJK}

\newcommand{\cH}{\mathcal{H}}
\newcommand{\cO}{\mathcal{O}}
\newcommand{\bx}{\bm{x}}

\begin{document}

\title{The $O(N)$ Free-Scalar and Wilson-Fisher Conformal Field Theories on the Fuzzy Sphere}

\author{Wenhan Guo}
\affiliation{C. N. Yang Institute for Theoretical Physics, Stony Brook University, Stony Brook, NY 11794-3840}
\author{Zheng Zhou (周正)}
\affiliation{Perimeter Institute for Theoretical Physics, Waterloo, Ontario N2L 2Y5, Canada}
\affiliation{Department of Physics and Astronomy, University of Waterloo, Waterloo, Ontario N2L 3G1, Canada}
\author{Tzu-Chieh Wei}
\affiliation{C. N. Yang Institute for Theoretical Physics, Stony Brook University, Stony Brook, NY 11794-3840}
\author{Yin-Chen He}
\affiliation{C. N. Yang Institute for Theoretical Physics, Stony Brook University, Stony Brook, NY 11794-3840}

\date{\today}

\begin{abstract}
    The fuzzy-sphere regularization is an emerging numerical and theoretical technique for studying conformal field theories (CFTs). In this paper, we apply it to the $O(N)$ vector model, one of the most prominent theories for critical behavior in three space-time dimensions. We construct a model that realizes the $O(N)$ Wilson-Fisher and free-scalar CFTs for general $N$. For $N=2,3,4$, we present numerical evidence including the operator spectra and correlation functions in agreement with conformal symmetry and conformal bootstrap results.
\end{abstract}

\begin{CJK*}{UTF8}{bkai}
\maketitle
\end{CJK*}

\section{\label{sec:Introduction}Introduction}

Conformal field theories (CFTs) are quantum or statistical field theories invariant under conformal transformations~\cite{Polyakov:1970xd,Belavin1984BPZ}. CFTs have many applications across different fields, ranging from condensed matter~\cite{Cardy:1996xt,Sachdev:2011fcc,Zinn-Justin:1989rgp} to string theory~\cite{Polchinski:1998rq}. In condensed matter physics, many continuous phase transitions are captured by CFTs. Two prominent examples are the 3D Ising transition and its generalization---the $O(N)$ Wilson-Fisher (WF) CFT~\cite{wilson1972critical}. The $O(N)$ WF theory describes the order-disorder transition between an $O(N)$ spontaneously broken phase and an $O(N)$ symmetric phase. It can emerge in various experimental setups~\cite{pelissetto2002critical}. For example, the exponents of $O(2)$ WF transition has been measured with high precision in the superfluid-to-normal transition of Helium-4 ($^4\mathrm{He}$)~\cite{cond-mat/0310163}, and the $O(3)$ WF transition has been observed in Heisenberg magnets~\cite{AlsNielson1976}. 

The $O(N)$ WF theory can be described by a quantum field theory (QFT) of $N$ interacting real scalar fields. An $O(N)$-symmetric quartic interaction term induces a renormalization-group (RG) flow from the ultraviolet (UV) fixed point of the $O(N)$ free-scalar theory to the interacting CFT fixed point in the infrared (IR). This formulation enables perturbative RG computations, such as the $4-\epsilon$ expansion originally proposed by Wilson and Fisher~\cite{wilson1972critical}. Over the decades, the $O(N)$ WF has been extensively studied using various theoretical methods (see a recent review~\cite{henriksson2023critical}), including perturbative computations ($4-\epsilon$~\cite{wilson1972critical,1705.06483,Schnetz:2016fhy,Schnetz:2022nsc,Henriksson:2025vyi} and large-$N$~\cite{Vasiliev:1981dg,hep-th/0306133} expansions) and non-perturbative calculations using conformal bootstrap~\cite{1504.07997,1603.04436,Chester:2020iyt,ChesterO2} or Monte Carlo simulations of lattice models~\cite{cond-mat/0605083,hasenbusch2000eliminating,Hasenbusch:2011zwv,Hasenbusch:2020pwj,Deng:2006tb,Hasenbusch:2005nx}. These different approaches have yielded accurate critical exponents. In particular, the conformal bootstrap~\cite{1805.04405,2311.15844} has achieved state-of-the-art precision for critical exponents and operator product expansion (OPE) coefficients~\cite{Chester:2020iyt,ChesterO2}.

The fuzzy-sphere regularization~\cite{zhu2023uncovering} has recently emerged as a versatile framework for studying 3D CFTs. The key idea is to investigate quantum many-body systems on a fuzzy (non-commutative) sphere~\cite{Madore:1991bw}. More concretely, one can take the highly degenerate lowest Landau level (LLL) as the single-particle ground state and construct on this basis interactions that realize quantum phase transitions~\cite{Ippoliti2018Half,Wang2021SO5WZW}. On a sphere, the spherical LLL is generated by a magnetic monopole~\cite{Sphere_LL_Haldane} at its center. One notable advantage of the fuzzy-sphere approach is that it naturally realizes 3D CFTs on the cylinder $S^2 \times \mathbb{R}$, a conformally flat space-time geometry, enabling the leverage of conformal symmetry---particularly the state-operator correspondence, where eigenstates of the critical Hamiltonian on the sphere are in one-to-one correspondence with CFT operators, and the energy gaps are proportional to their scaling dimensions. The fuzzy-sphere approach has significantly advanced our understanding of 3D CFTs. Beyond the 3D Ising CFT~\cite{zhu2023uncovering,hu2023operator,han2023conformal,hu2024entropic,hofmann2023quantum,fardelli2024constructing,fan2024note,voinea2024regularizing,lauchli2025exact,2507.20005,2510.09482}, a wealth of information has been extracted regarding its conformal line defects~\cite{hu2024solving,zhou2024g,Cuomo2024} and boundaries~\cite{zhou2025studying,dedushenko2024ising}. The method has been applied to a variety of other theories, including the $SO(5)$~\cite{zhou2024so5,chen2024emergent} and $O(4)$~\cite{Yang2025Jul} deconfined quantum phase transitions (DQCP), the Heisenberg bilayer~\cite{Han2023Dec}, symplectic gauge theories~\cite{zhou2025new}, the 3-state Potts transition~\cite{yang2025microscopic}, the Lee-Yang non-unitary CFT~\cite{fan2025simulating,cruz2025yang,miro2025flowing}, the free real scalar~\cite{he2025free,Taylor2025}, the free Majorana fermion~\cite{Zhou2025Sep}, the confinement transition of the $\nu=1/2$ bosonic Laughlin state~\cite{Zhou2025Jul}, and the quantum Hall transition between the Halperin $(221)$ state and bosonic Pfaffian~\cite{Voinea2025}.

Given that the $O(N)$ WF CFT has been extensively investigated using various well-established methods, it serves as an ideal testbed for the fuzzy-sphere approach. More importantly, the fuzzy sphere paves the way for studying many aspects of the $O(N)$ WF that may be difficult to access through other approaches. For instance, it enables a more direct study of primary operators with specific quantum numbers (e.g., $O(N)$ pseudo-scalar, pseudo-vector, etc.), OPE coefficients, and the $F$-function~\cite{Jafferis:2011zi,Casini2012}; it allows the access to multi-scalar CFTs (such as cubic, tetrahedral, and orthogonal bifundamental theories)~\cite{Aharony:1973zz,Toledano1985,Osborn:2017ucf,Zia:1975ha,Rong:2023owx,Hasenbusch:2022zur,Kousvos:2025ext,Rong:2023xhz}; it also lays the groundwork for investigating various conformal line defects (such as pinning-field~\cite{Cuomo:2021kfm,ParisenToldin:2016szc} and spin-impurity~\cite{Cuomo2022Spin,Komargodski:2025jbu,Kaj2007} defects) and conformal boundaries~\cite{Metlitski:2020cqy,Toldin:2021kun,Gliozzi:2015qsa,Hasenbusch_2011} in the $O(N)$ WF. 

In this paper, we introduce two classes of $O(N)$ models that can respectively
realize the $O(N)$ WF and $O(N)$ free-scalar CFTs at arbitrary $N$.
As a generalization of the Ising~\cite{zhu2023uncovering}, free scalar~\cite{he2025free}, and $O(2)$ WF~\cite{Lauchli_XY} models, we consider $(N+1)$ flavors of fermions $\psi_{0,1,\dots,N}$ on the fuzzy sphere. Among them, $\psi_0$ is an $O(N)$ singlet and $\psi_{1,\dots,N}$ transform as an $O(N)$ vector. We set the total filling $\nu=1$. The WF CFT describes a transition between a paramagnetic phase, realized when $\psi_0$ is fully filled, and an $O(N)$ SSB phase, which has one flavor of $O(N)$ superposed with the $0$-flavor. Turning to the $O(N)$ free-scalar theory, it corresponds to the pseudo-Goldstone phase of a $PO(N+1)$ SSB. Numerically, we verify that our models realize the $O(N)$ WF and free-scalar CFTs for $N=2,3,4$. We report the parameters for the respective CFTs and demonstrate that the operator spectra and two-point correlators exhibit conformal symmetry and are consistent with the established conformal-bootstrap and Monte Carlo results. We discuss several quantities that may
not have a direct CFT interpretation. The results imply that the CFTs become rather semi-classical once regularized on the fuzzy sphere. 

The structure of the remainder of this paper is as follows. In Sec.~\ref{sec:HamiltonianAnsatz}, we present fuzzy-sphere models for the $O(N)$ WF and free-scalar CFTs. In Sec.~\ref{sec:Result}, we present our numerical results: We determine the parameters for each CFT, calculate the operator spectrum and conformal correlators, and discuss seome non-CFT quantities. We conclude our study in Sec.~\ref{sec:Discussion}. In the Appendices, we provide useful formulae related to the fuzzy-sphere formalism, our numerical conventions and technical details, and a comprehensive walkthrough on how the numerical computation was carried out.

\section{Model Hamiltonian \label{sec:HamiltonianAnsatz}}

\begin{figure}[b]
    \centering
    \includegraphics[width=\linewidth]{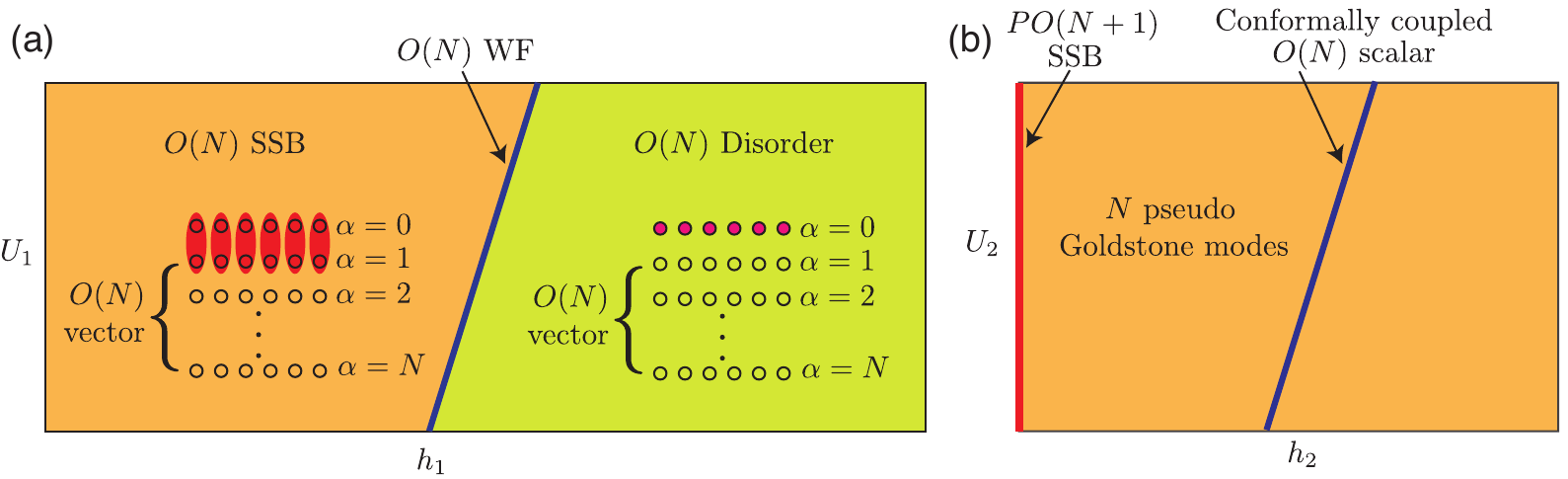}
    \caption{\label{fig:phasediagram} A schematic phase diagram of the $O(N)$ WF and $O(N)$ free-scalar model.}
\end{figure}

The philosophy of the fuzzy-sphere regularization is similar to that of condensed matter lattice model realizations of CFTs: one realizes the CFT of interest as the universality class of a critical point (phase) in an interacting quantum-mechanical model. Concretely, it realizes a 3D CFT as the low-energy theory of interacting fermions in the LLL on a two-sphere threaded by a magnetic monopole. We refer readers unfamiliar with the method to the original proposal~\cite{zhu2023uncovering}, the recent review~\cite{he2026fuzzy}, and the FuzzifiED package~\cite{zhou2025fuzzified} for further background, physical motivation, and numerical details.

We consider $(N+1)$-flavor fermions $\psi^\dag_\alpha(\bm x)$ moving on a sphere and subject to a monopole with flux $4\pi s$. $\psi_0(\bm x)$ is the $O(N)$ singlet, and $\psi_{1,\cdots, N}(\bm x)$ form an $O(N)$ vector. In the following, we use the convention that Greek indices $\alpha,\beta=0,1,\cdots, N$ label all flavors including the singlet, while Latin indices $a,b=1,\dots,N$ label only the $O(N)$-vector components. The monopole creates quantized Landau levels for fermions, and we project the system to the LLL~\cite{Sphere_LL_Haldane}, which consists of $N_\text{orb}=2s+1$ orbitals forming a spin-$s$ representation of $SO(3)$ sphere rotation. The Landau-level gap is a free parameter, independent of the interaction scale, which we may take to infinity; in this limit the projection to the LLL becomes exact, and since the gap is a UV scale it does not affect the IR universality. The model is defined by fermion operators $c_{m,\alpha}$ living on the $2s+1$ Landau orbitals, $m=-s,-s+1,\cdots, s$ and $\alpha=0,1,\cdots, N$. The operators $c_{m,\alpha}$'s follow the anticommutation relation of canonical fermions, $\{ c_{m_1,\alpha}, c_{m_2,\beta}\}=\delta_{m_1,m_2}\delta_{\alpha\beta}$. 
The fermion field $\psi_\alpha(\bm x)$ in real space is defined as
\begin{equation}
    \psi^\dagger_\alpha(\bm x) =\frac{1}{\sqrt{2s+1}}\sum_{m=-s}^s c^\dagger_{m,\alpha} Y^{(s)}_{s,m}(\bm x),
    \end{equation}
$Y^{(s)}_{s,m}(\bm x)$ is the wavefunction of LLL, called monopole harmonics (omitting the $1/\sqrt{4\pi}$ factor),
 \begin{multline}\label{eq:monopoleHarmonics}
Y^{(s)}_{s,m}(\bm x) = \sqrt{\frac{(2s+1)!}{(s+m)!(s-m)!}} e^{im\varphi} \\\times\cos^{s+m}\left(\frac{\theta}{2}\right)\sin^{s-m}\left(\frac{\theta}{2}\right).
 \end{multline}
 We can define composite fields using these fermion fields, 
 \begin{align}
 n_{\alpha\beta}(\bm x) & = \psi^\dag_\alpha(\bm x) \psi_\beta(\bm x) + \psi^\dag_\beta(\bm x) \psi_\alpha(\bm x),  \label{eq:nalphabeta} \\ 
 \omega_{\alpha\beta}(\bm x) & = i(\psi^\dag_\alpha(\bm x) \psi_\beta(\bm x) - \psi^\dag_\beta(\bm x) \psi_\alpha(\bm x) ).
 \end{align}
The fields $n_{\alpha\beta}(\bm{x})$ and $\omega_{\alpha\beta}(\bm{x})$ are Hermitian, i.e., $(n_{\alpha\beta}(\bm{x}))^\dag = n_{\alpha\beta}(\bm{x})$ and $(\omega_{\alpha\beta}(\bm{x}))^\dag = \omega_{\alpha\beta}(\bm{x})$. Under the exchange of $\alpha$ and $\beta$, $n_{\alpha\beta}(\bm{x})$ is symmetric, while $\omega_{\alpha\beta}(\bm{x})$ is antisymmetric. These composite fields can be viewed as the analogs of quantum fields in conventional QFT, or of lattice spin operators in condensed matter lattice models. Especially, we can define an $O(N)$-vector spin field $V_{a=1,\cdots,N}(\bm x)$,
\begin{equation}
V_a(\bm x) = n_{0a}(\bm x) = n_{a0}(\bm x),
\end{equation}
which is the order parameter of the $O(N)$ spontaneous symmetry breaking. 

We use these composite fields to define the Hamiltonian in the continuous real space, as elaborated below, but the actual computation is performed in the Landau orbital basis in terms of the $c_{m,\alpha}$ fermions. In the fermion representation $c_{m,\alpha}$, the number $2s + 1$ plays the role of the system size; more precisely, the radius of the sphere is $R = \sqrt{2s + 1}\,l_B$. The magnetic length $l_B$ serves as an analog of the lattice spacing and plays the role of a short-distance cutoff. Different choices of $l_B$ do not affect the infrared physics.
For ease of notation, we set $l_B=1$ and therefore omit it in the remainder of the paper, so that $N_{\text{orb}} = 2s + 1 = R^2$. The continuum limit is realized as $s \rightarrow \infty$. We consider a situation where the fermion particle number is conserved, $\sum_{m,\alpha} c_{m,\alpha}^\dag c_{m,\alpha} = 2s + 1$, i.e., one fermion per Landau orbital.

The Hamiltonian for $O(N)$ WF is given by $H_1=\int d^2\bm x\,\cH_1(\bm x)$, where
\begin{equation} \label{eq:WFAnsatz}
\cH_{1}(\bm x) = (n_{tot}(\bm x))^2  - U_1 \sum_{a=1}^N V_{a}(\bm x) \nabla^2 V_a(\bm x) + h_1 n_{00}(\bm x).
\end{equation}
The first term is the square of the total particle density, $n_{\text{tot}}(\bm{x}) = \frac{1}{2}\sum_{\alpha=0}^N n_{\alpha\alpha}(\bm{x})$, which is $SU(N+1)$ symmetric. From the symmetry perspective, this term is not strictly necessary; however, it has been empirically observed, in a variety of fuzzy-sphere models~\cite{zhu2023uncovering,zhou2025new,he2025free,Zhou2025Sep,Zhou2025Jul}, that such a term helps reduce finite-size effects. Physically, it suppresses local charge fluctuations, which we believe underlies the reduced finite-size effects. Being $SU(N+1)$-symmetric, this term does not alter the universality class, and a complete understanding of why it is so effective remains an interesting open question. The second term breaks the $SU(N+1)$ symmetry down to $O(N)$ and can be viewed as the analog of a nearest-neighbor spin-spin interaction in a lattice model. The last term is a chemical potential that controls particle occupation and serves as the tuning parameter for the order-disorder transition. When $h_1$ is small (e.g., $h_1 \ll -1$), the particles prefer to occupy the $\alpha = 0$ flavor, so the ground state is
\begin{equation}\label{eq:disorder}
 |\psi_{\text{$O(N)$ disorder}}\rangle = \prod_m c^\dag_{m,0}  |0\rangle,   
\end{equation}
which is $O(N)$ symmetric, i.e., disordered. On the other hand, a larger $h_1$ excites particles to the last $N$ flavors, $a = 1, \cdots, N$, and the spin-spin interaction further favors an $O(N)$ spontaneous symmetry breaking (SSB) state, i.e.,
\begin{equation}\label{eq:SSB}
    |\psi_{\text{$O(N)$ SSB}}(\hat{\bm e})\rangle = \prod_m \frac{\sum_{a=1}^Nc^\dag_{m,a}\hat{e}_a + c^\dag_{m,0} }{\sqrt{2}} |0\rangle,   
\end{equation}
where $\hat{\bm e}$ is a vector on the $S^{N-1}$ unit sphere that captures the $O(N)$ SSB. The phase transition between these two phases is reached by tuning $h_1$ and is described by the $O(N)$ WF CFT. A schematic phase diagram for arbitrary $N$ is shown in Fig.~\ref{fig:phasediagram}. We remark that the $N = 1$ model indeed reduces to the original fuzzy-sphere Ising model~\cite{zhu2023uncovering}, and the disorder state in Eq.~\eqref{eq:disorder} and SSB state in Eq.~\eqref{eq:SSB} correspond to all spins pointing to the $z$ or $x$ direction, respectively. 

 It is worth noting that one can also write down another symmetry-breaking state, 
  \begin{equation}\label{eq:PONSSB}
 |\psi_{\text{$PO(N)$ SSB}}\rangle = \prod_m \left( \sum_{a=1}^N\hat{e}_a c^\dagger_{m,a} \right)|0\rangle,   
\end{equation}
This state only breaks $PO(N)=O(N)/\{\pm I\}$ symmetry~\footnote{For odd $N$, the $\mathbb{Z}_2$ projected is the improper-$\mathbb{Z}_2$ and by $PO(N)$ we mean $SO(N)$.}---the $\mathbb Z_2$ center $c_{m,a} \mapsto - c_{m,a}$ of $O(N)$ remains unbroken. Another way to understand it is to evaluate the $O(N)$ SSB order parameter $V_a(\bm x)$, which gives $ \langle\psi_{\text{$PO(N)$ SSB}}| V_a(\bm x)\hat{e}_a |\psi_{\text{$PO(N)$ SSB}}\rangle=0$ but $ \langle\psi_{\text{$O(N)$ SSB}}| V_a(\bm x)\hat{e}_a |\psi_{\text{$O(N)$ SSB}}\rangle=1$. We also remark that the ground state manifold of $O(N)$ SSB is $S^{N-1}$, while the ground state manifold of $PO(N)$ SSB is $RP^{N-1}$.

The $O(N)$ free-scalar CFT is a UV fixed point of the $O(N)$ vector model. It can still describe an order-disorder transition, but with more relevant operators. Thus, the free scalar can appear at a multi-critical point, and a natural idea is to search for such a point by introducing additional tuning parameters into the $O(N)$ WF Hamiltonian in Eq.~\eqref{eq:WFAnsatz}. This idea has been pursued for the ($O(1)$) real scalar~\cite{Taylor2025}. Here, we follow another approach, originally as an alternative proposal for the real scalar~\cite{he2025free}, which requires fewer tuning parameters and, more importantly, maintains a closer connection with the standard QFT. We consider instead  $H_2=\int d^2\bm x\,\cH_2(\bm x)$, where
\begin{multline}\label{eq:GaussianAnsatz}
\cH_2(\bm x) = \left(n_{tot}(\bm x)\right)^2  \\+ U_2 \sum_{0\le\alpha<\beta\le N} \omega_{\alpha\beta}(\bm x) \nabla^2 \omega_{\alpha\beta}(\bm x) + \frac{h_2}{R^2} n_{00}(\bm x).
\end{multline} 
Here we take the radius of the sphere as $R=\sqrt{N_\text{orb}}=\sqrt{2s+1}$.

When $h_2 = 0$, the Hamiltonian has an $O(N+1)$ flavor symmetry, with the fermions $\psi_{\alpha=0,1,\cdots,N}$ forming an $O(N+1)$ vector. We also note that the $\mathbb{Z}_2$ center of the $O(N+1)$ symmetry, $\psi_\alpha(\bm{x}) \mapsto -\psi_\alpha(\bm{x})$, acts trivially on all bosonic operators of the model. Thus, for bosonic operators, the global symmetry is $PO(N+1) = O(N+1)/\{\pm I\}$. The ground state at $h_2=0$ is an SSB state, so the effective theory is a non-linear sigma model (NLSM). Since the NLSM is a theory of bosonic fields, its global symmetry should be $PO(N+1)$ rather than $O(N+1)$, and the target space of NLSM is $RP^N = S^N / \mathbb{Z}_2$ instead of $S^N$. As in the $S^N$ NLSM, the $RP^N$ NLSM contains $N$ gapless Goldstone modes. It is worth noting that the symmetry breaking order is the anti-ferromagnetic order such that the Goldstone modes have linear dispersions. 

A finite $h_2$ breaks $PO(N+1)$ down to $O(N)$ rather than $PO(N)$ because $V_a(\bm{x}) = n_{0a}(\bm{x})$ transforms as an $O(N)$ vector on which the center $\mathbb{Z}_2$ acts non-trivially. This term also pins the original SSB pattern to the $0$-direction, and the $N$ gapless Goldstone modes become pseudo-Goldstone modes with mass $m \sim \sqrt{h_2}/R$. These can be described by an $O(N)$ scalar field $\phi_{a=1,\cdots,N}$ with the effective theory
\begin{align}
\mathcal{L} &= \frac{1}{2} \partial_\mu \phi_a \, \partial^\mu \phi_a  
+ \frac{h_2}{2f} \frac{\phi_a \phi_a}{R^2} \nonumber \\
&\quad + \frac{1}{2f^2} (\phi_a \partial_\mu \phi_a)^2 
+ \frac{h_2}{8f^3} \frac{(\phi_a \phi_a)^2}{R^2} + \cdots,
\end{align}
where $f$ is the stiffness associated with the $PO(N+1)$ spontaneous symmetry breaking.
The first two terms correspond to the standard free-scalar theory. In particular, the scalar mass scales as the inverse sphere radius $1/R$, making it a marginal perturbation at the free-scalar fixed point. On the sphere one must fine-tune this marginal coupling $\phi_a \phi_a / R^2$ in order to obtain a conformally coupled free scalar (see,  e.g.,~Ref.~\cite{BirrellDavies:1982}). The effective theory also contains interaction terms, but these are irrelevant, especially the quartic term $(\phi_a \phi_a)^2$, because it is further suppressed by a factor of $1/R^2$. Therefore, microscopically it suffices to fine-tune $h_2$ in order to reach the conformally coupled free-scalar fixed point in our model. A schematic phase diagram for arbitrary $N$ is shown in Fig.~\ref{fig:phasediagram}.

\section{Numerical Results\label{sec:Result}}

In this section, we present the numerical evidence that the fuzzy-sphere models \eqref{eq:WFAnsatz} and \eqref{eq:GaussianAnsatz} realize respectively $O(N)$ WF and free-scalar CFTs at $N=2,3,4$, including the conformal operator spectra and the conformal two-point correlation functions.

We simulate the fuzzy-sphere models using exact diagonalization~\cite{zhou2025fuzzified}. The symmetry of the models includes
\begin{itemize}[nosep]
    \item a $U(1)$ fermion number conservation that decouples at the critical points, 
    \item the $SO(3)$ sphere rotation symmetry (note that in this model, the spatial parity, which is usually realized as the particle-hole symmetry in the fuzzy-sphere models, is absent), and 
    \item the $O(N)$ global symmetry, which includes the connected part $SO(N)$ as a subgroup of the $SU(N)$ complex rotation of the $1,\dots,N$ flavors, as well as the improper $\mathbb{Z}_2$ that acts as adding a $(-1)$ factor to one of the flavors (see Appendix \ref{app:ChoosingRepresentativeStates}).
\end{itemize}
In the ED calculation, we implement the $U(1)$ and discrete subgroups of the global symmetry, which block-diagonalizes the Hamiltonian and segments the Hilbert space. The maximal system size we can reach is $N_\text{orb}=11$ for $O(2)$, $10$ for $O(3)$ and $9$ for $O(4)$. 

\begin{figure}[b]
    \centering
    \includegraphics[width=0.9\linewidth]{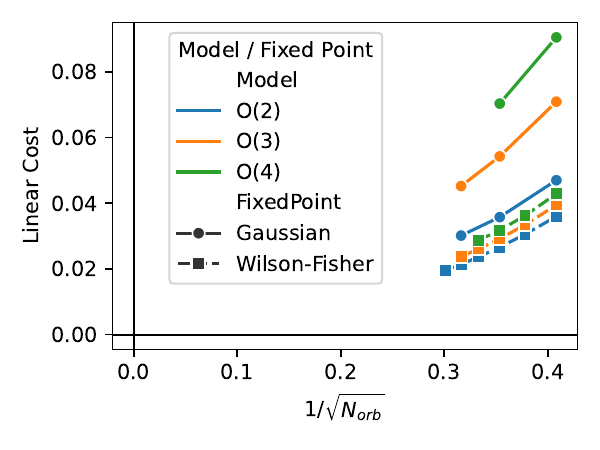}
    \caption{The minimal spectral conformal cost function for the $O(N)$ ($N=2,3,4$) WF and free-scalar CFTs as a function of the inverse linear system size $1/\sqrt{N_\text{orb}}$. The parameters for the miximal size are listed in Table~\ref{tab:params}. The free-scalar CFTs have a worse conformal cost at finite system sizes and an even-odd effect of $N_\text{orb}$, so we focus on even $N_\text{orb}$.}
    \label{fig:lin_cost}
\end{figure}

With the emergent conformal symmetry, the operator spectrum can be extracted from the state-operator correspondence; namely, each eigenstate $|\cO\rangle$ of the critical quantum Hamiltonian has a one-to-one mapping to the local operator $\cO$ of the CFT. The scaling dimension of the operator is proportional to the excited energy of the eigenstate
\begin{equation}
    \Delta_\cO = \frac{v}{R} (E_{\cO} - E_0),
    \label{eq:FSCal}
\end{equation}
where the $E_0$ is the ground state energy, and $v$ is a model-dependent constant to be determined that can be interpreted as the ``speed of light."  The state and the operator carry the same quantum numbers under the global symmetry, so that the Lorentz spin can be determined by measuring the total angular momentum $\langle\cO|L^2|\cO\rangle=\ell_\cO(\ell_\cO+1)$ of the state, and the $O(N)$ representation can be determined by the quadratic Casimir $C_2$ (see Appendix \ref{app:ChoosingRepresentativeStates}).

\subsection{Convergence of Conformal Cost Function}

To locate the parameters in the ansatzes (\ref{eq:WFAnsatz}) and (\ref{eq:GaussianAnsatz}) that realize respectively the WF and free-scalar CFTs, we compare the spectrum of the fuzzy-sphere model with the operator spectrum of the desired CFT. 

These CFTs have two conserved operators: the symmetry current $J^\mu_{[ab]}$ at $\Delta_J=2$ with Lorentz spin-$1$ in the $O(N)$ anti-symmetric rank-$2$ tensor representation, and the stress tensor $T^{\mu\nu}$ with Lorentz spin-$2$ in the $O(N)$ singlet representation. Moreover, the scaling operators organize into conformal multiplets, each consisting of a primary operator and its descendants at integer spacing formed by acting spatial derivatives. Specifically, for the $O(N)$ free-scalar and WF CFTs, the lowest-lying primaries include the vector field $\phi_a$, the lowest singlet $S=\phi^2$, and the $O(N)$ rank-2 tracesless symmetric rank-2 tensor $t_{(ab)}=\phi_a\phi_b-\delta_{ab}\phi^2/N$ where $(ab)$ and $[ab]$ denote respectively the symmetrization and anti-symmetrization of the indices, and we will omit the substraction of trace hereafter. For the free scalars, they have exact scaling dimensions $\Delta_\phi=1/2$ and $\Delta_t=\Delta_S=1$; these scaling dimensions get corrected in the WF CFTs and can be most accurately calculated from numerical conformal bootstrap~\cite{1504.07997,1603.04436,Chester:2020iyt,ChesterO2}.

\begin{table}[t]
\centering
\setlength{\tabcolsep}{5pt}
\renewcommand{\arraystretch}{1.1}
\begin{tabular}{lcc}
\hline
Model & $U_1$ & $h_1$ \\
\hline
$O(2)$ WF & $0.270$ & $-0.084$  \\
$O(3)$ WF & $0.264$ & $-0.096$  \\
$O(4)$ WF & $0.255$ & $-0.106$  \\
\hline
Model & $U_2$ & $h_2$ \\
\hline
$O(2)$ Gaussian & $0.148$ & $-0.070$ \\
$O(3)$ Gaussian & $0.140$ & $-0.070$ \\
$O(4)$ Gaussian & $0.132$ & $-0.072$ \\
\hline
\end{tabular}

\caption{The parameters for the $O(N)$ ($N=2,3,4$) WF and free-scalar CFTs in the Hamiltonian ansatzes \eqref{eq:WFAnsatz} and \eqref{eq:GaussianAnsatz}, optimized at the maximal accessible system size. }
\label{tab:params}
\end{table}

\begin{figure*}[htp]
    \centering
    \includegraphics[width=0.95\linewidth]{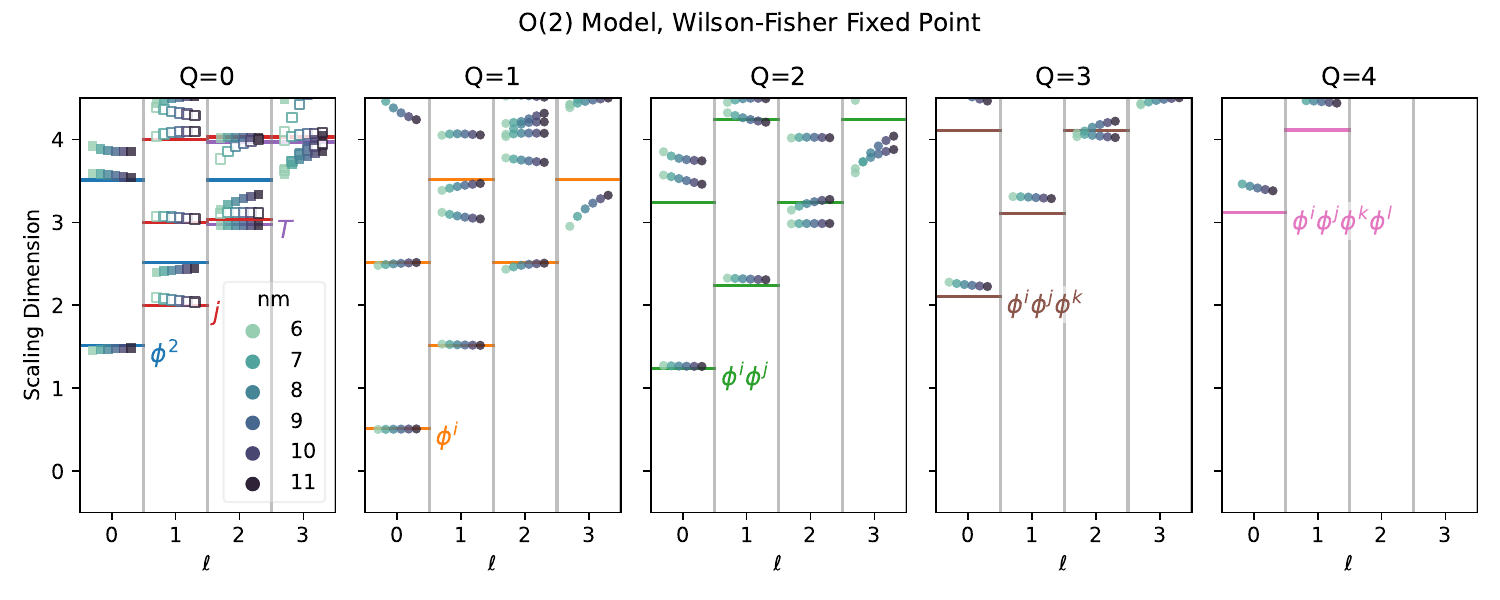}\\
    \includegraphics[width=0.95\linewidth]{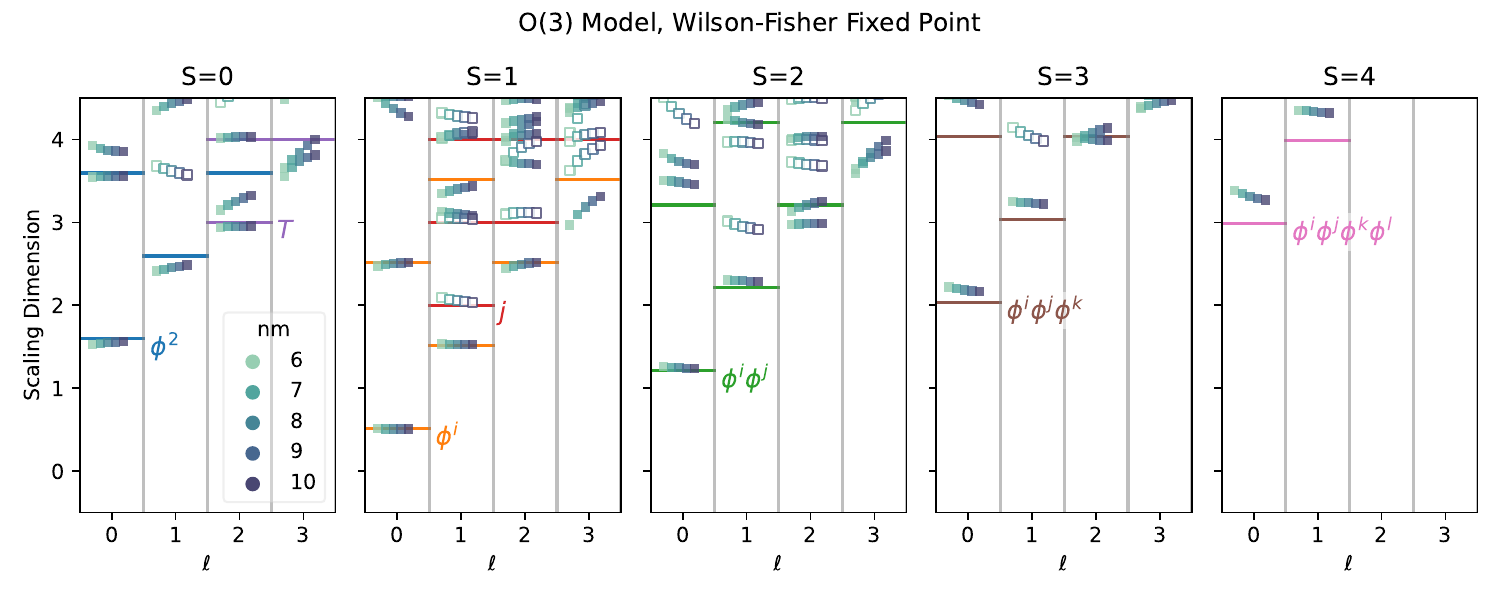}\\
    \includegraphics[width=0.95\linewidth]{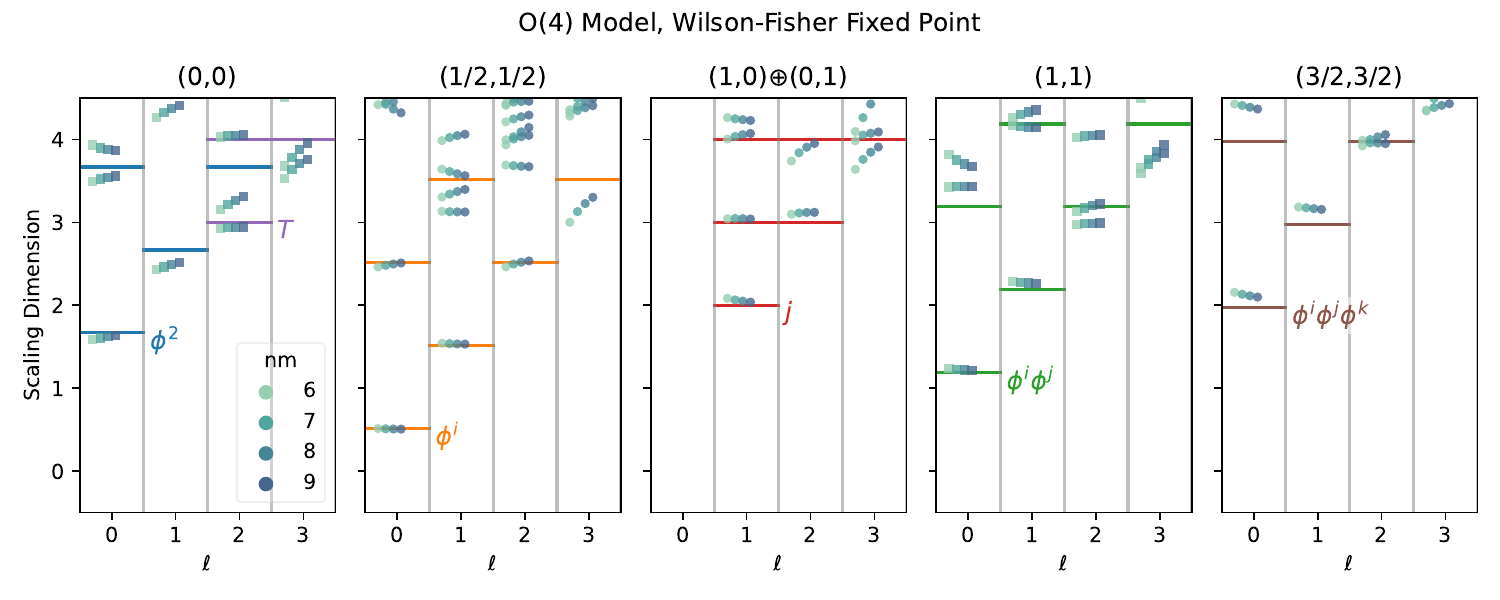}
    \caption{Operator spectrum of the $O(N)$ ($N=2,3,4$) WF CFT from state-operator correspondence at different system sizes at their respective optimal conformal point up to $\Delta\lesssim4,\ell\leq 3$. The filled, empty squares and the circles denote respectively operators with improper $\mathbb{Z}_2$-even, odd, and no definite improper $\mathbb{Z}_2$. The bars are bootstrap~\cite{ChesterO2,Chester:2020iyt} and Monte Carlo~\cite{hasenbusch2000eliminating,Hasenbusch:2011zwv} results and the states in the same multiplet are labeled by the same color.}
    \label{fig:scdim_WF}
\end{figure*}

\begin{figure*}[ht]
    \centering
    \includegraphics[width=0.49\linewidth]{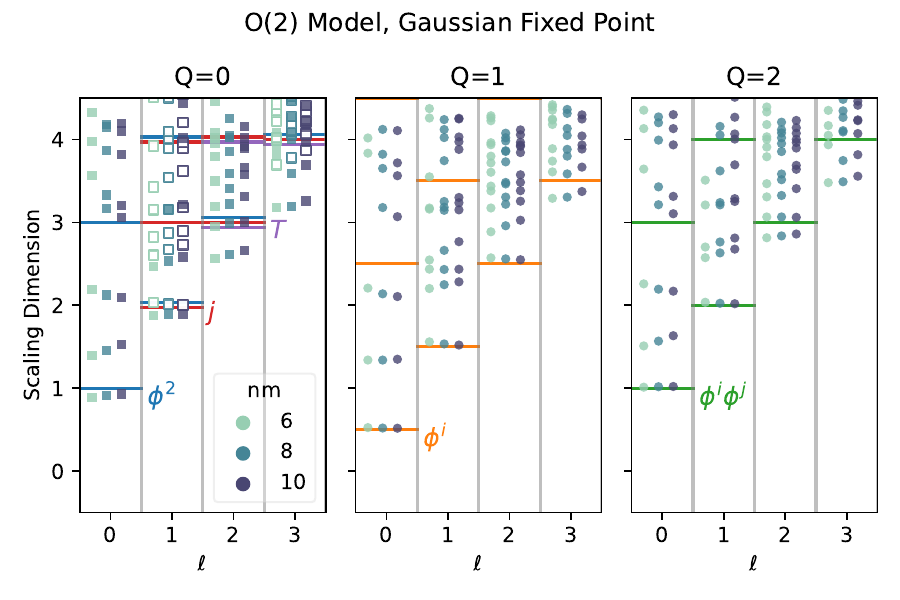}
    \includegraphics[width=0.49\linewidth]{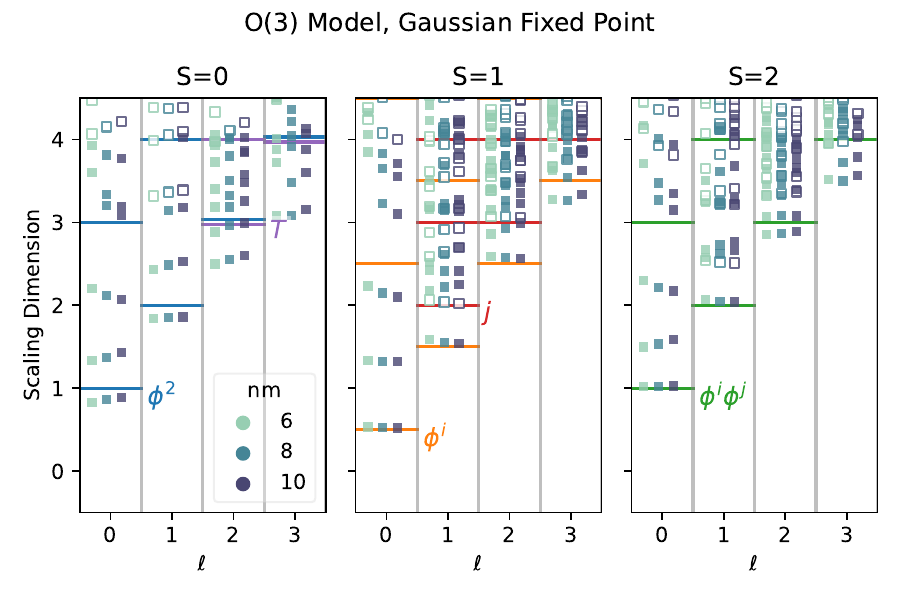}\\
    \includegraphics[width=0.63\linewidth]{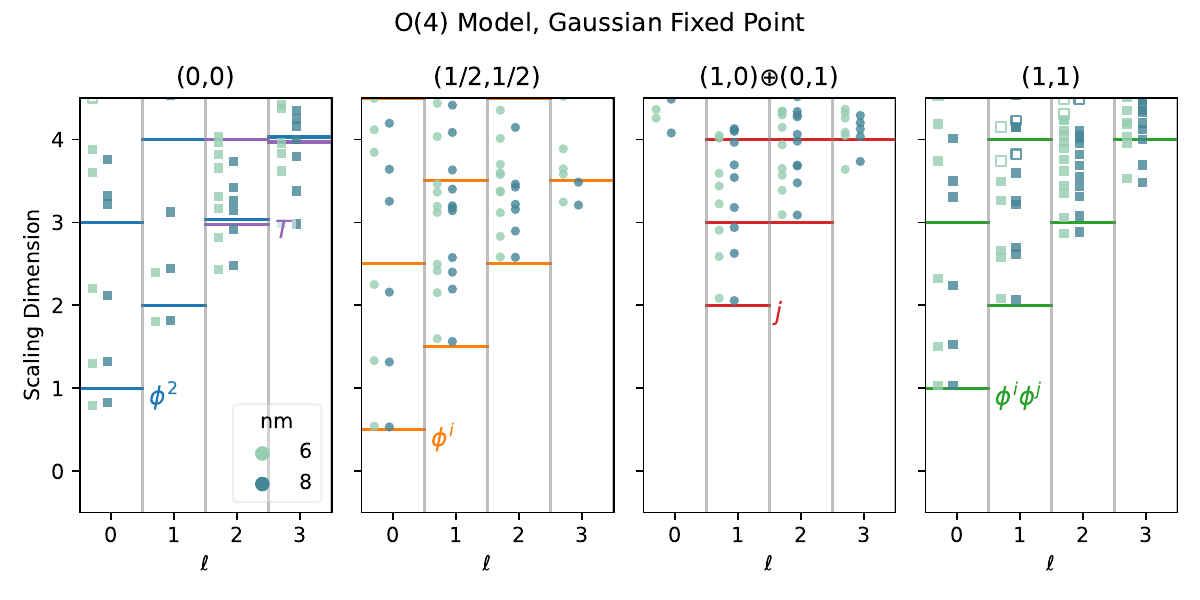}
    \caption{Operator spectrum of the $O(N)$ ($N=2,3,4$) free-scalar CFT from state-operator correspondence at different system sizes at their respective optimal conformal point up to $\Delta\lesssim4,\ell\leq 3$ and the rank-2 tensor representations. The filled, empty squares and the circles denote respectively operators with improper $\mathbb{Z}_2$-even, odd and no definite improper $\mathbb{Z}_2$. The bars are theoretical results and the colors label the multiplets. }
    \label{fig:scdim_Gaussian}
\end{figure*}

Numerically, to compare the spectra with the expected CFTs, we define a spectral conformal cost function as the weighted root-mean square of the difference between the scaling dimensions from the fuzzy sphere and their reference values
\begin{equation}
    Q_{\rm cost} = \sqrt{\sum_\cO W_\cO\left(\Delta_\cO^{\text{(FS)}}-\Delta_\cO^{\text{(ref)}}\right)^2}.
\end{equation}
Here $\Delta_\cO^{\text{(FS)}}$ is calculated from Eq.~\eqref{eq:FSCal}, and reference values $\Delta_\cO^{\text{(ref)}}$ are taken as the exact results for the free scalars and from bootstrap~\cite{ChesterO2,Chester:2020iyt} and Monte Carlo~\cite{hasenbusch2000eliminating,Hasenbusch:2011zwv} for the WF CFTs. The operators we include in the sum are $\phi,S,t$, their lowest descendants up to $\Delta=3$, the conserved current $J^\mu$, and the stress tensor $T^{\mu\nu}$. These operators are identified via their Lorentz spins and the $O(N)$ representations. The weights $W_\cO$ are taken to enhance sensitivity to the low-lying operators; they sum to one. For details, see Appendix~\ref{app:cost_function}.

For $N = 2, 3, 4$, we optimize the cost function for the parameters $(U_1, h_1)$ and $(U_2, h_2)$ for the WF and free theory, respectively, along with the velocity $v$ for each system size $N_{\text{orb}}$. For all the models we consider, the minima of the cost function decrease with system size $N_{\text{orb}}$ and display a trend converging toward zero in the thermodynamic limit $N_{\text{orb}}$ (Fig.~\ref{fig:lin_cost}). For the free-scalar models, the cost function additionally shows an even-odd effect in $N_{\rm orb}$ at finite size; this is a finite-size artifact, and the even and odd $N_{\rm orb}$ sequences extrapolate to the same IR CFT. As a sanity check against overfitting, we find that the cost function is rather insensitive as one moves along the $(U, h)$ critical line near the optimal parameters. Moreover, the optimal parameters remain stable across system sizes $N_{\text{orb}}$. These results support that the Hamiltonians (\ref{eq:WFAnsatz}) and (\ref{eq:GaussianAnsatz}) realize the WF and free-scalar CFTs, respectively. In the rest of paper, we will present the results computed at the optimal parameters, shown in Table~\ref{tab:params}. 

\begin{figure*}[ht]
    \centering
    \includegraphics[width=0.8\linewidth]{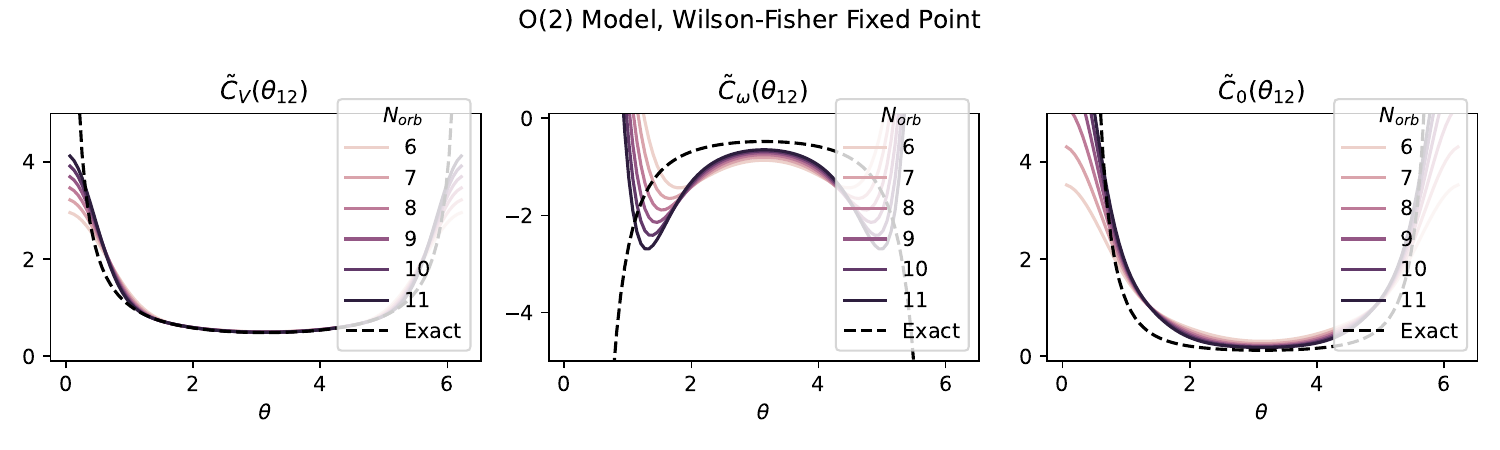}
    \includegraphics[width=0.8\linewidth]{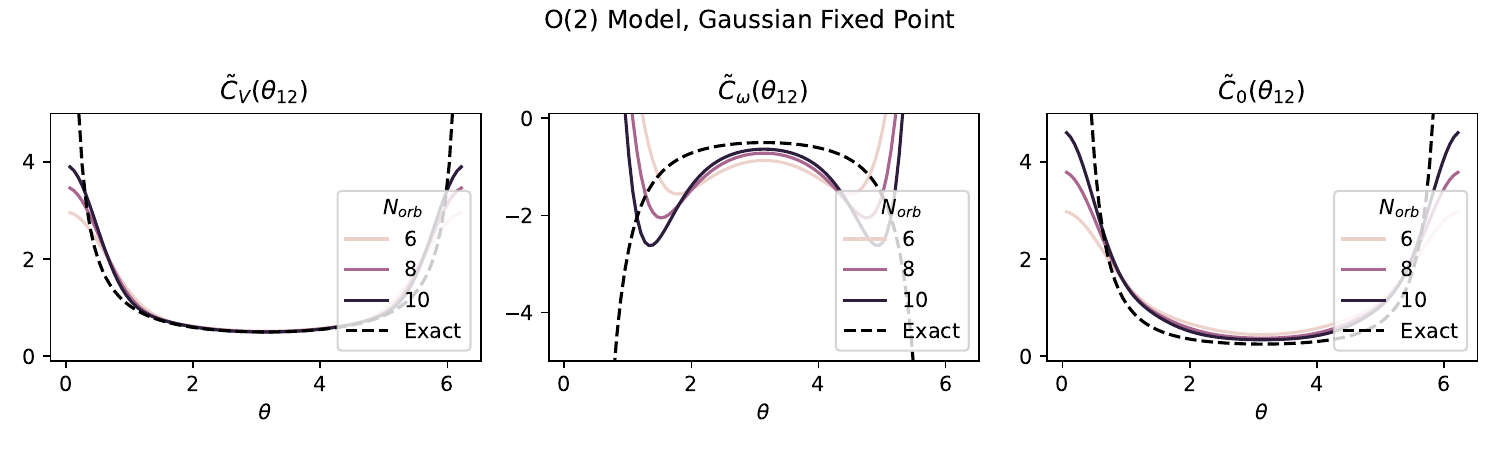}
    \includegraphics[width=0.8\linewidth]{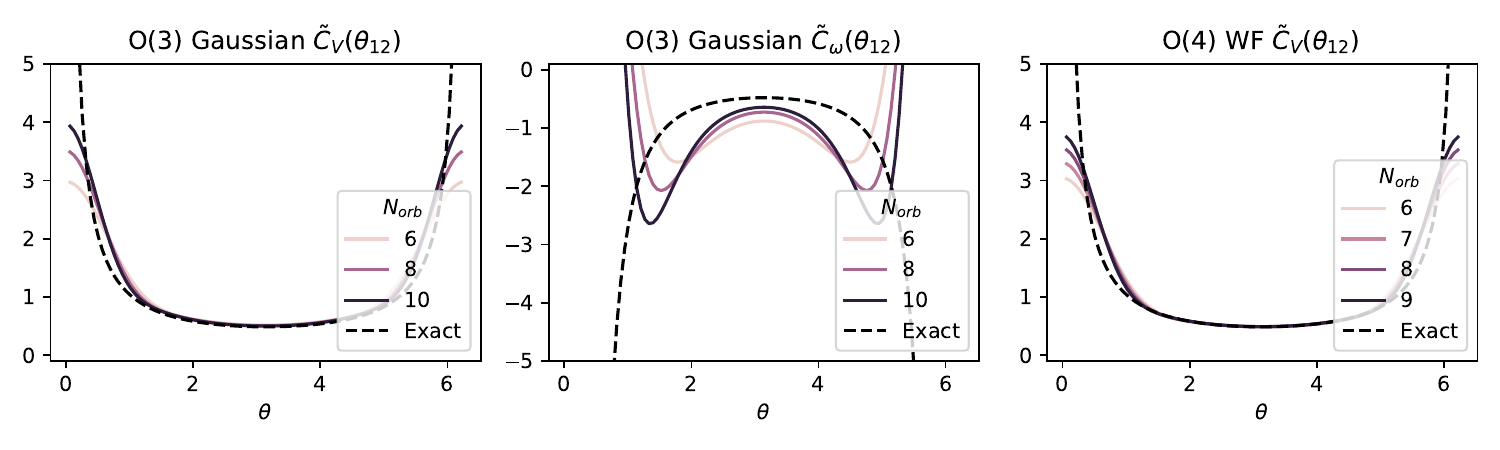}
    \caption{Dimensionless two-point correlation functions of UV fields $V_a(\bm x)$, $\omega_a(\bm x)$, and $n_{00}(\bm x)$ in the $O(2)$ fuzzy-sphere model for the WF and free scalar. In the IR, these correlators should converge to the CFT correlators of $\phi_a(\bm x)$,  $\pi_a(\bm x)=\partial_\tau \phi_a(\bm x)$ and $S(\bm x)$, shown as the dashed black line. }
    \label{fig:corr_O2_WF}
\end{figure*}

\subsection{Scaling Dimensions and Conformal Tower}

Having located the optimal parameters, we take a closer look at the spectra and show that the operators that are not included in the cost function also agree with the $O(N)$ WF and free-scalar CFTs. We plot the low-lying spectrum for the $O(N)$ ($N=2,3,4$) WF and free-scalar CFTs extracted from state-operator correspondence at different system sizes in Figs.~\ref{fig:scdim_WF} and \ref{fig:scdim_Gaussian}.

For the WF CFTs (Fig.~\ref{fig:scdim_WF}), our numerical results reproduce the scaling dimensions of the lowest operators ($\phi,t,S$) and the conserved currents ($J,T$) with a relative discrepancy with their expected values typically around 2\%. In addition, all their expected descendant can be found, and all the lowest levels can be organized into conformal multiplets. For a Lorentz singlet $\cO$, its descendants can be written in the form of $\partial^{\nu_1}\dots\partial^{\nu_j}\Box^n\cO$ ($n,j\geq0$) with spin-$j$ and scaling dimension $\Delta+2n+j$; for spinning primary $\cO^{\mu_1\dots\mu_\ell}$, its descendants can be written either as $\partial^{\nu_1}\dots\partial^{\nu_j}\Box^n\partial_{\mu_1}\dots\partial_{\mu_i}\cO^{\mu_1\dots\mu_\ell}$ ($0\le i\le \ell, n,j\ge 0$) with spin-$(\ell-i+j)$ and scaling dimension $\Delta+2n+i+j$ or as $\epsilon_{\mu_1\rho\sigma}\partial^\rho\partial^{\nu_1}\dots\partial^{\nu_j}(\partial^2)^n\partial_{\mu_2}\dots\partial_{\mu_i}\cO^{\mu_1\dots\mu_\ell}$ ($1\le i\le \ell, n,j\ge 0$) with spin-$(\ell-i+j+1)$ and scaling dimension $\Delta+2n+i+j$; for conserved currents only $i=0$ descendants exist due to the conservation $\partial_{\mu_1}\cO^{\mu_1\dots\mu_\ell}=0$. 

For the free-scalar CFTs, most operators converge toward the expected scaling dimensions with increasing $N_\text{orb}$, while some deviations are still visible. This is expected as the free-scalar CFTs exhibit a higher degree of degeneracy in its low-energy spectrum. These degenerate states are more sensitive to perturbations from irrelevant operators, leading to visible splittings, a phenomena known as ``level repulsion'' in the conformal perturbation theory~\cite{lauchli2025exact}.

\subsection{Correlation Functions}

Apart from the operator spectrum, another evidence for the conformal symmetry comes from the two point correlation functions of the local operators. For a scalar primary $\cO$, the conformal symmetry restricts the form of its two-point function on the sphere 
\begin{align}
    C_\cO(x_{12})&=\langle \cO(\bx_1)\cO(\bx_2)\rangle=\nonumber\\
    &=\frac{1}{x_{12}^{2\Delta_\cO}}=\frac{1}{R^{2\Delta_\cO}\left(2- 2\cos\theta_{12}\right)^{\Delta_\cO}}\label{eq:CFT_2p_fn},
\end{align}
where $\bx_{1,2}$ are two points on the sphere with distance $x_{12}=|\bx_1-\bx_2|$ and angular distance $\theta_{12}$. 

These CFT local operators can be realized by the composite fields on the fuzzy sphere. At the conformal point, each composite field can be expressed as a linear combination of the CFT scaling operators with the same quantum numbers under $O(N)$ symmetry and time-reversal (which acts as the complex conjugation). The simplest ones are $O(N)$ vectors $V_a=n_{0a}$, $\omega_a=\omega_{0a}$, respectively even and odd under time reversal, and the singlet $n_{00}$. To the leading order,
\begin{align}
    V_a(\bx)&=c_\phi\phi_a(\bx)+\dots\nonumber\\
    \omega_a(\bx)&=c_\pi\pi_a(\bx)+\dots\nonumber\\
    n_{00}(\bx)&=c_0\mathbb{I}+c_SS(\bx)+\dots
\end{align}
where the conjugate momentum fields $\pi_a=\partial_\tau\phi_a$ is a descendant of $\phi_a$. The dots represents the subleading contributions from higher primary or descendant operators in the same representation, such as $\phi^2\phi_a$ or $\partial^\mu\phi_a$ to $V_a$ and $\phi^4$ to $n_{00}$. 

The UV-dependent coefficients $c_\cO$ can be cancelled by defining the dimensionless correlators
\begin{align} \label{eq:UV_corr1}
    \tilde C_V(\theta_{12})&= \frac{\langle0|V_a(\bx_1) V_a(\bx_2)|0\rangle}
    {|\langle\phi_a|V_a(\bx)|0\rangle|^2},\\ \label{eq:UV_corr2}
    \tilde C_\omega(\theta_{12})&= \frac{\langle0|\omega_a(\bx_1) \omega_a(\bx_2)|0\rangle}
    {|\langle\phi_a|\omega_a(\bx)|0\rangle|^2}, \\
    \tilde C_0(\theta_{12})&= \frac{\langle0|n_{00}(\bx_1) n_{00}(\bx_2)|0\rangle-\langle0|n_{00}(\bx) |0\rangle^2} 
    {|\langle S|n_{00}(\bx)|0\rangle|^2} \label{eq:UV_corr3},
\end{align}
where $\bx$ is an arbitrary point on the sphere. To the leading order, these dimensionless correlators depend only on the angular distance and the scaling dimensions of the corresponding CFT operators 
\begin{align}
    \tilde C_V(\theta_{12})&=\frac{1}{(2-2 \cos\theta)^{\Delta_\phi}} ,\label{eq:IR_corr1} \\
    \tilde C_\omega(\theta_{12})&=\frac{-2/\Delta_\phi}{(2-2 \cos\theta)^{\Delta_\phi+1}}, \label{eq:IR_corr2}\\
    \tilde C_0(\theta_{12})&=\frac{1}{(2-2 \cos\theta)^{\Delta_S}} \label{eq:IR_corr3}.
\end{align}
With their respective $\Delta_\phi$ and $\Delta_S$, these relations hold for both the WF and free-scalar CFT. 
Numerically, we calculate the dimensionless correlators~\eqref{eq:UV_corr1}--\eqref{eq:UV_corr3} for different system sizes for $O(N)$ ($N=2,3,4$) WF and free-scalar CFTs and compare them with the CFT results. The results are plotted in Fig.~\ref{fig:corr_O2_WF}. As $N_\text{orb}$ increases, they converge to the CFT two-point functions~\eqref{eq:IR_corr1}--\eqref{eq:IR_corr3}, which provides further evidence for conformal symmetry. 

\subsection{Non-CFT Observables}

In this section, we discuss several quantities that may not have a direct CFT interpretation. Nevertheless, these quantities exhibit rather peculiar behavior compared to conventional expectations, and thus may have interesting implications for the question of why the fuzzy-sphere approach has such a small finite-size effect. 

\begin{figure}
    \centering
    \includegraphics[width=0.49\linewidth]{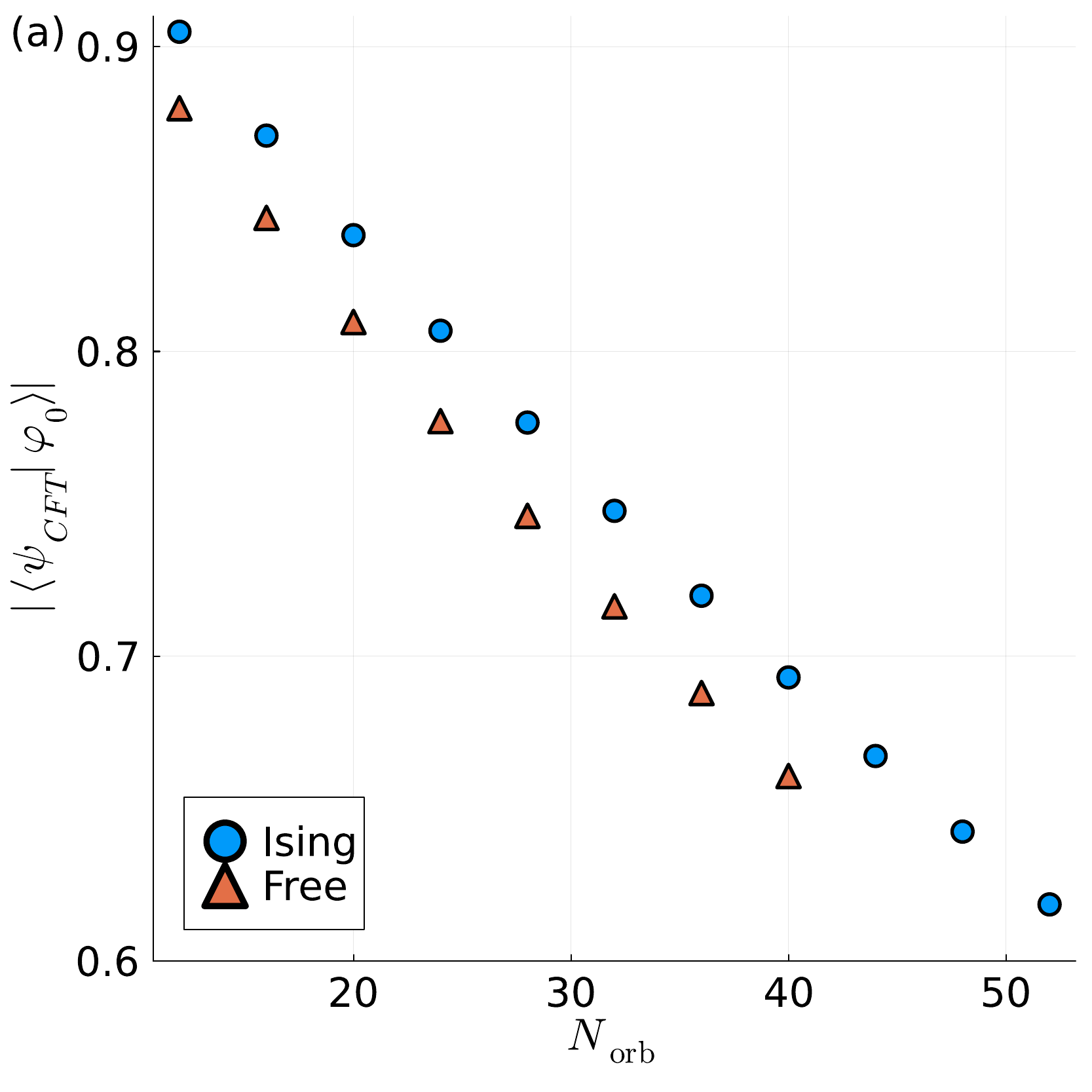}
    \includegraphics[width=0.49\linewidth]{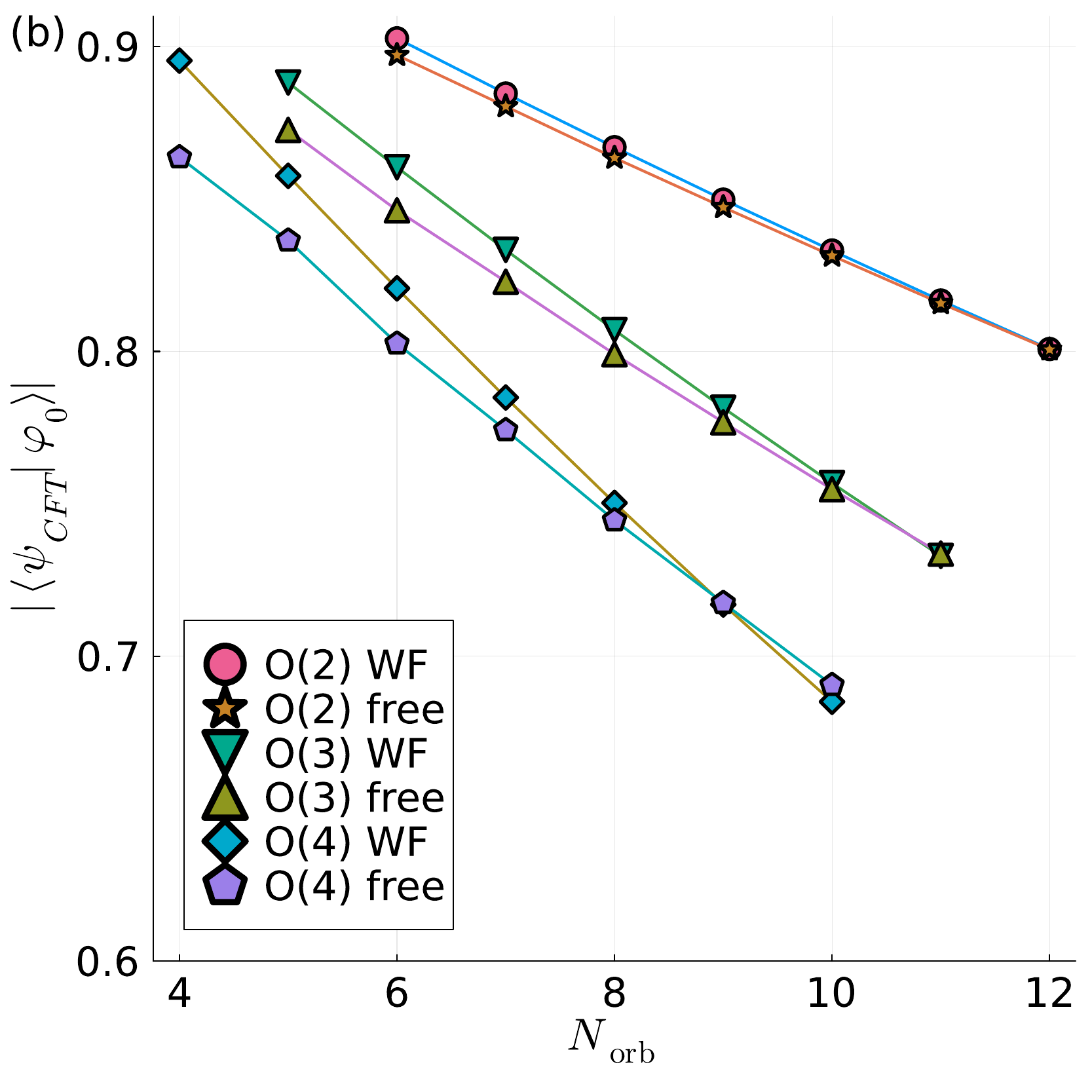}
    \caption{The wavefunction overlap between the trivial product state and 3D CFT ground state: (a)  Ising and free real ($O(1)$) scalar; (b) $O(N)$ WF and free scalar.}
    \label{fig:overlap}
\end{figure}

The most unexpected observation is that the CFT ground state is close to a trivial product state, $|\varphi_0\rangle = \prod_m c^\dag_{m,0}|0\rangle$.
This state is an $O(N)$ singlet, and in the case of the Ising CFT ($N=1$), it corresponds to the fully polarized state. 
Fig.~\ref{fig:overlap}(a) shows the wavefunction overlap $|\langle \psi_{\mathrm{CFT}}|\varphi_0\rangle|$ for the 3D Ising CFT and the free scalar~\cite{he2025free}, which is unusually large; for example, the overlap exceeds $0.8$ for the Ising CFT at $N_{\mathrm{orb}} = 24$~\footnote{The boundary CFT central charge can be extracted from the wavefunction overlap~\cite{zhou2025studying}.}. 
Therefore, the CFT ground state can be viewed as a state that includes small fluctuations around the trivial product state, which can be well described by a wavefunction ansatz in a form similar to a coherent state~\cite{he2025free}.
The $O(N=2,3,4)$ WF and the free scalar also exhibit similar behavior, as shown in Fig.~\ref{fig:overlap}(b).

\begin{figure}
    \centering
    \includegraphics[width=0.49\linewidth]{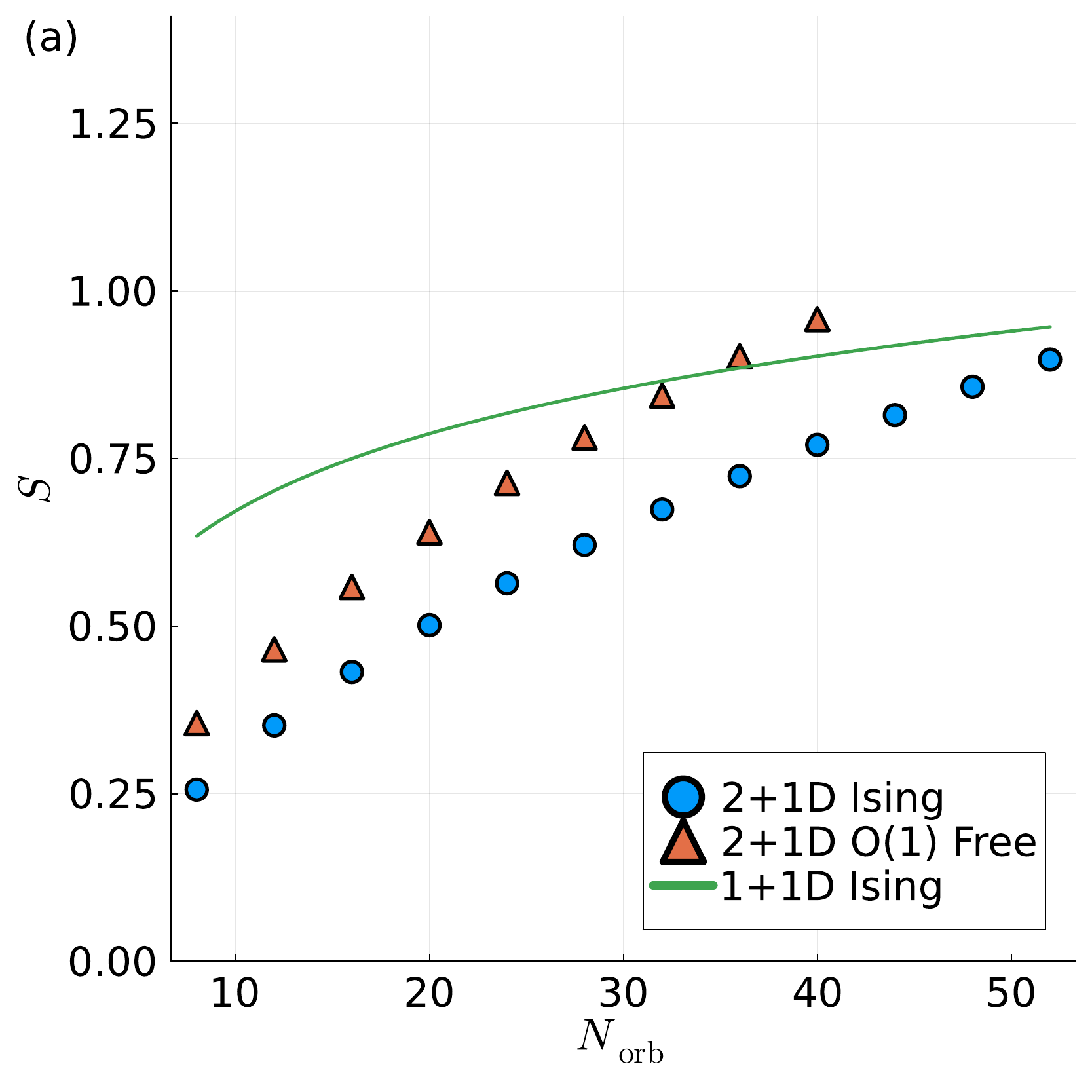}
        \includegraphics[width=0.49\linewidth]{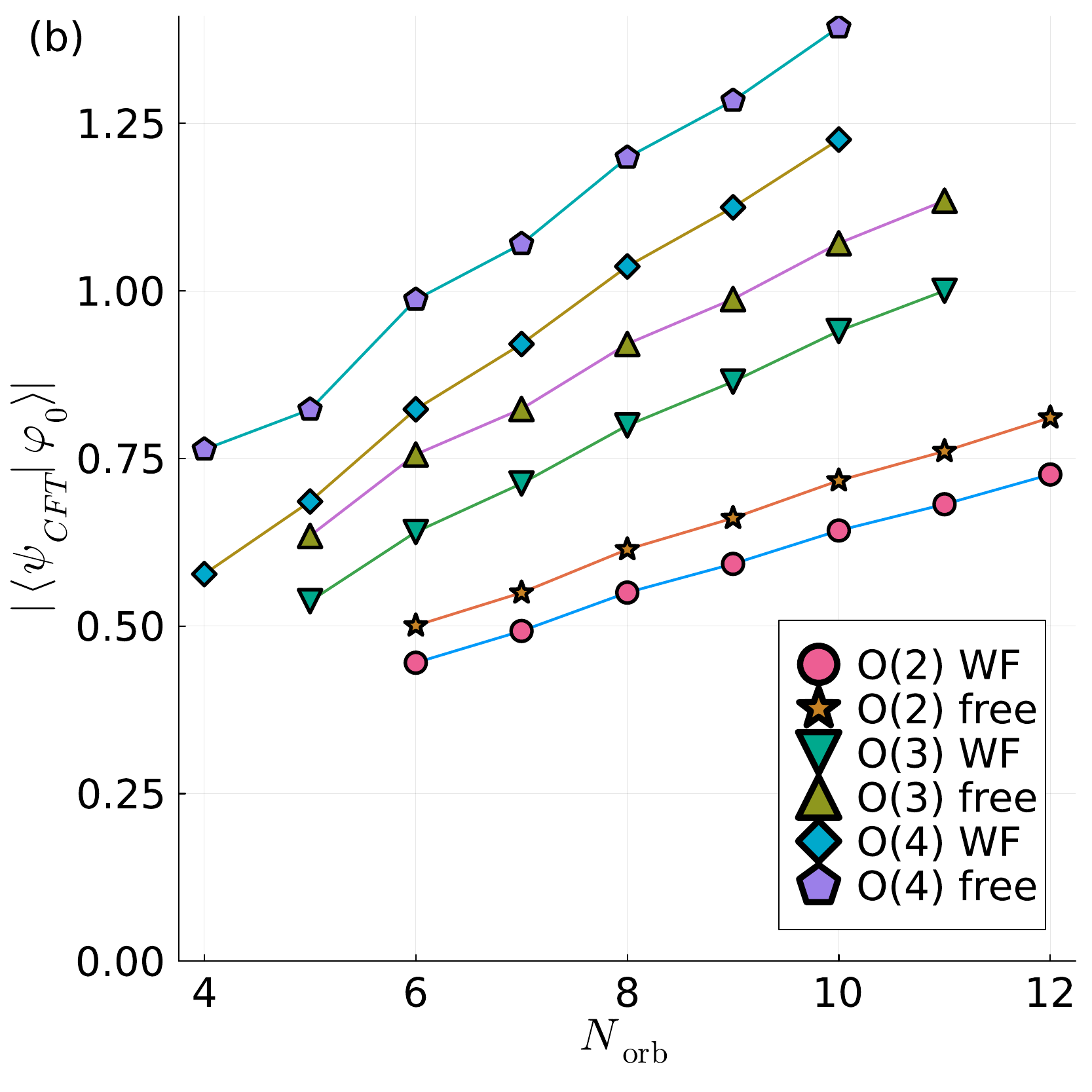}
    \caption{The entanglement entropy of  (a)  3D Ising, free real ($O(1)$) scalar  ground state, as well as 2D Ising CFT in the $(1+1)$D transverse-field Ising model; (b) $O(N)$ WF and free scalar.}
    \label{fig:EE}
\end{figure}

A consequence of the CFT ground state on the fuzzy sphere being close to a trivial product state is its small entanglement entropy, as shown in Fig.~\ref{fig:EE}. Practically, this also facilitates tensor-network-based methods such as DMRG to study 3D CFTs on the fuzzy sphere. To illustrate how small the entanglement entropy is, we compare it to that of the $(1+1)$D transverse-field Ising model (TFIM), which is also well known for its small entanglement entropy. For the TFIM, the number of orbitals is taken as the length of the spin chain.
Fig.~\ref{fig:EE}(a) shows the entanglement entropy for both the fuzzy sphere and the $(1+1)$D TFIM, where in each case we consider the entanglement entropy obtained by bipartitioning the system into two halves, resulting in maximal entanglement. 
We note that the 3D CFT on the fuzzy sphere is expected to satisfy the entanglement area law $S = \alpha \sqrt{N_{\mathrm{orb}}} + \beta$, while the $(1+1)$D CFT follows the entanglement scaling $S = \frac{c}{3} \ln\left(\frac{N_{\mathrm{orb}}}{\pi}\right) + S_0$. 
Therefore, at sufficiently large $N_{\mathrm{orb}}$, the $3D$ CFT should have a larger entanglement entropy. 
However, somewhat surprisingly, for the largest system size studied, $N_{\mathrm{orb}} = 52$, the entanglement entropy of the 3D Ising CFT is smaller than that of the $(1+1)$D TFIM. The $O(N=2,3,4)$ WF and the free scalar also have small entanglement entropy, as shown in Fig.~\ref{fig:EE}(b).

These observations defy the common lore that CFTs are strongly fluctuating and highly entangled; instead, they become rather semi-classical once regularized on the fuzzy sphere. 
We should remark that by semi-classical, we do not mean weakly coupled, as one would sometimes associate with the WF CFT due to the fact that $\Delta_\phi$ is close to the value of a free scalar field.
Indeed, for the quantities we have studied, the WF CFT behaves more classically than the free scalar: its wavefunction is closer to the trivial product state and it exhibits smaller entanglement entropy. 
Among all the theories considered, the Ising CFT appears to be the most classical one, even when compared to the free real scalar. 
It would be interesting to understand whether there is a deeper physical reason behind these observations.

\section{\label{sec:Discussion}Conclusion}

We propose two classes of fuzzy-sphere models for realizing $O(N)$ CFTs, namely the $O(N)$ WF and free-scalar theories. Using exact diagonalization, we verify that the operator spectrum and two-point correlators of our fuzzy-sphere models for $N=2,3,4$ agree with CFT expectations. We also show that the CFT ground states of these models are rather semi-classical, in the sense that they are close to a trivial product state and exhibit relatively small entanglement entropy. Among these theories, both WF and free-scalar CFTs, the Ising CFT is the most semi-classical. These observations defy the common lore that CFTs exhibit strong fluctuations and warrant further exploration.

Our fuzzy-sphere $O(N)$ model may help to explore several important aspects of the $O(N)$ WF CFT that are difficult to access using other approaches. For example, it would be interesting to compute the OPE of the symmetry current, $J^\mu \times J^\nu$, which provides direct access to the conductivity of quantum critical points that can be measured experimentally~\cite{Chowdhury:2012km,Katz2014}. It is also intriguing to study conformal line defects and boundaries of the $O(N)$ WF theory. One particularly compelling direction is to investigate spin impurities~\cite{Cuomo2022Spin,Komargodski:2025jbu,Kaj2007}, analogous to studying a Kondo impurity in a CFT. Another interesting problem is to analyze the RG flow between the free-scalar CFT and the $O(N)$ WF CFT. Finally, we remark that it would also be valuable to explore whether the fuzzy-sphere $O(N)$ model can be treated perturbatively using a large-$N$ expansion.

\begin{acknowledgments}

We would like to Davide Gaiotto and Chong Wang for fruitful discussions.
This work was partially supported by the US National Science Foundation under the Grant No. PHY 2310614.
Research at Perimeter Institute is supported in part by the Government of Canada through the Department of Innovation, Science and Industry Canada and by the Province of Ontario through the Ministry of Colleges and Universities. Z.~Z.~acknowledges support from the Natural Sciences and Engineering Research Council of Canada (NSERC) through Discovery Grants.
The numerical calculations are done using the package FuzzifiED~\cite{zhou2025fuzzified}.

\paragraph*{Note added.} While preparing this manuscript, we became aware of a parallel study~\cite{Dey2025} that investigates the $\mathrm{O}(3)$ WF CFT with a similar fuzzy-sphere model.

\end{acknowledgments}

\appendix
\section{Decomposition into the Orbital Space\label{app:Yslm}}

\newcommand{\rd}{\mathrm{d}}
\newcommand{\bnh}{\hat{\mathbf{n}}}
\newcommand{\bzh}{\hat{\mathbf{z}}}
\newcommand{\br}{\mathbf{r}}
\newcommand{\cP}{\mathcal{P}}
\newcommand{\cR}{\mathcal{R}}
\newcommand{\cZ}{\mathcal{Z}}
\newcommand{\BI}{\mathbb{I}}
\newcommand{\BR}{\mathbb{R}}
\newcommand{\BZ}{\mathbb{Z}}
\newcommand{\Yb}{\bar{Y}}
\newcommand{\Int}{\text{int}}
\allowdisplaybreaks

We project the system onto the LLL by writing the fermion operators in terms of the annihilation operators of the LLL orbitals
\begin{equation}
    \psi^\dagger_\alpha(\br)=\frac{1}{R}\sum_{m=-s}^s Y^{(s)}_{sm}(\bnh)c^\dagger_{m\alpha},
\end{equation}
where $Y_{lm}^{(s)}(\bnh)$ is the monopole harmonics, $\bnh$ is the unit vector of the point on the sphere specified by angular coordinates $\theta$ and $\phi$, $c^{(\dagger)}_{m\alpha}$ annihilates/creates an electron with $L^z$-quantum number $m$ at the $f$-th flavour of the lowest Landau level.

Like the fermion operator, the density operator, \textit{i.~e.}~local fermion bilinear
\begin{equation}
    n_M(\br)=\psi_{f'}^\dagger(\br)M_{f'f}\psi_\alpha(\br).
    \label{eq:den_def}
\end{equation}
can also be expressed in the orbital space
\begin{equation}
    n_M(\br)=\sum_{lm}Y_{lm}(\bnh)n_{M,lm}.
    \label{eq:den_decomp}
\end{equation}
Conversely,
\begin{widetext}
\begin{align}
    n_{M,lm}&=\frac{1}{R^2}\int\rd^2\br\,\Yb_{lm}(\bnh)n_M(\br)\nonumber\\
    &=\int\rd^2\bnh\,\Yb_{lm}(\bnh)\left(\frac{1}{R}\sum_{m_1}Y^{(s)}_{sm_1}(\bnh)c^\dagger_{m_1f_1}\right)M_{f_1f_2}\left(\frac{1}{R}\sum_{m_2}\Yb^{(s)}_{sm_2}(\bnh)c_{m_1f_2}\right)\nonumber\\
    &=\frac{1}{R^2}\sum_{m_1m_2}c^\dagger_{m_1f_1}M_{f_1f_2}c_{m_2f_2}\int\rd^2\bnh\,\Yb_{lm}(\bnh)Y^{(s)}_{sm_1}(\bnh)\Yb^{(s)}_{sm_2}(\bnh)\nonumber\\
    &=\frac{1}{R^2}\sum_{m_1}c^\dagger_{m_1f_1}M_{f_1f_2}c_{m_1-m,f_2}(-1)^{s-m_1}(2s+1)\sqrt{\frac{2l+1}{4\pi}}\begin{pmatrix}l&s&s\\-m&m_1&-m_1+m\end{pmatrix}\begin{pmatrix}l&s&s\\0&-s&s\end{pmatrix}.
    \label{eq:den_mod}
\end{align}
Here, we have used the properties of the monopole spherical harmonics~\cite{Wu:1976ge}
\begin{align}
    \Yb_{lm}^{s}&=(-1)^{s+m}Y_{l,-m}^{(-s)}\nonumber\\
    \int\rd^2\bnh\,Y_{lm}^{(s)}\Yb_{lm}^{(s)}&=\delta_{ll'}\delta_{mm'}\nonumber\\
    \int\rd^2\bnh\,Y_{l_1m_1}^{(s_1)}Y_{l_2m_2}^{(s_2)}Y_{l_3m_3}^{(s_3)}&=\sqrt{\frac{(2l_1+1)(2l_2+1)(2l_3+1)}{4\pi}}\begin{pmatrix}l_1&l_2&l_3\\m_1&m_2&m_3\end{pmatrix}\begin{pmatrix}l_1&l_2&l_3\\-s_1&-s_2&-s_3\end{pmatrix},
\end{align}
\end{widetext}
where $\left(\begin{smallmatrix}\bullet&\bullet&\bullet\\\bullet&\bullet&\bullet\end{smallmatrix}\right)$ is the $3j$-symbol, and 
$Y_{lm}^{(0)}=Y_{lm}$ is the common spherical harmonics. In this way, we have fully expressed the density operator in terms of the operators in the orbital space $c^{(\dagger)}_{m\alpha}$.

The most straightforward way to construct an interaction term is to add a density-density interaction with a potential function
\begin{equation}
    H_\Int=\int\rd^2\br_1\,\rd^2\br_2\,U(|\br_1-\br_2|)n_M(\br_1)n_M(\br_2).
\end{equation}
The interacting potentials can be expanded in terms of the Legendre polynomials
\begin{align}
    U(|\br_{12}|)&=\sum_l\tilde{U}_lP_l(\cos\theta_{12})\nonumber\\
    &=\sum_{lm}\frac{4\pi\tilde{U}_l}{2l+1}\Yb_{lm}(\bnh_1)Y_{lm}(\bnh_2),
\end{align}
where $\br_{12}=\br_1-\br_2$ and $|\br_{12}|=2R\sin\theta_{12}/2$. Conversely
\begin{equation}
    \tilde{U}_l=\int\sin\theta_{12}\rd\theta_{12}\,\frac{2l+1}{2}U(|\br_{12}|)P_l(\cos\theta_{12}).
\end{equation}
Specifically, for local and super-local interactions
\begin{align}
    U(|\br_{12}|)&=g_0\delta(\br_{12}),&\tilde{U}_l&=\frac{g_0}{R^2}(2l+1)\nonumber\\
    U(|\br_{12}|)&=g_1\nabla^2\delta(\br_{12}),&\tilde{U}_l&=-\frac{g_1}{R^4}l(l+1)(2l+1).
\end{align}
Here we make use of the conversion relations 
\begin{equation*}
    \delta(\br)=\frac{1}{R^2}\delta(\bnh),\qquad \nabla^2_\br=\frac{1}{R^2}\nabla^2_{\bnh},
\end{equation*}
for $\br=R\bnh$. By expanding the density operators into the orbital space and completing the integrals,
\begin{equation}
    H_\Int=\sum_{lm}\frac{4\pi R^4\tilde{U}_l}{2l+1}n^\dagger_{M,lm}n_{M,lm}.
\end{equation}

One can similarly convert the correlation functions in the real space and orbital space 
\begin{align}
    C(\theta_{12})&=\langle n_M(\br_1)n_M(\br_2)\rangle\nonumber\\
    &=\sum_{lml'm'}\Yb_{lm}(\br_1)Y_{l'm'}(\br_2)\langle n^\dagger_{lm} n_{l'm'}\rangle\nonumber\\
    &=\sum_{lml'm'}\Yb_{lm}(\br_1)Y_{l'm'}(\br_2)\delta_{ll'}\delta_{mm'}\langle n^\dagger_{l0} n_{l0}\rangle\nonumber\\
    &=\sum_{l}\frac{2l+1}{4\pi}P_l(\cos\theta_{12})\langle n^\dagger_{l0} n_{l0}\rangle
\end{align}
The third line makes use of the $SO(3)$ rotation symmetry. Hence, to obtain $C(\theta_{12})$ as a continuous function of $\theta_{12}$, one only needs to evaluate $\langle n^\dagger_{l0} n_{l0}\rangle$.

\section{Cost Function\label{app:cost_function}}

In this section, we explain the definition of the spectral cost function, which evaluates how close a fuzzy-sphere spectrum is to the reference CFT. We determine the optimal data point and the light speed $v$ by minimising the cost function.

We choose a collection of states and list their energies 
\begin{multline}
    E = [E_T, E_j, E_\phi, E_s, E_t, E_{\partial \phi} - E_\phi, \\E_{\partial s} - E_s, E_{\partial t} - E_t, E_{\partial\partial \phi} - E_\phi],
\end{multline}
where $S=\phi^i\phi^i$ is the singlet (scalar), $\phi^i$ is the flavor vector, $t^{ij}=\phi^{(i}\phi^{j)}$ is the symmetric traceless tensor. We compare them with the list of reference target scaling dimensions from engineering dimension of free scalars or conformal bootstrap
\begin{equation}
\Delta = [3, 2, \Delta_\phi^{\text{(ref)}}, \Delta_s^{\text{(ref)}}, \Delta_t^{\text{(ref)}}, 1, 1, 1, 2].
\end{equation}
For an ideal CFT, these two vectors are proportional
\begin{equation}
\Delta_i = \alpha E_i,
\end{equation}
where $\alpha=v/R$ is a factor to be determined. We define the cost function as:
\begin{equation}
Q_{\rm cost} = \left[\frac{1}{\sum_i W_i} \sum_i W_i (\Delta_i - \alpha E_i)^2\right]^{1/2}.
\end{equation}
We choose the weights as the reference scaling dimensions to the power $-2$ to reduce the contribution from higher-lying states:
\begin{equation}
W = [3, 2, \Delta_\phi^{\text{ref}}, \Delta_s^{\text{ref}}, \Delta_t^{\text{ref}}, \Delta_{\partial \phi}^{\text{ref}}, \Delta_{\partial s}^{\text{ref}}, \Delta_{\partial t}^{\text{ref}}, \Delta_{\partial\partial \phi}^{\text{ref}}]^{-2}.
\end{equation}
The optimal rescaling factor $\alpha$ is computed as:
\begin{equation}
\alpha = \frac{\sum_i W_i E_i \Delta_i}{\sum_i W_i E_i^2}.
\end{equation}

\section{Basis Transformation\label{app:complexfieid_basis}}

\newcommand{\ct}{\tilde{c}}
\newcommand{\Vt}{\tilde{V}}

To facilitate the exact diagonalisation calculation, we need to perform a basis transformation. Taking the $O(2)$ model as an example, the $O(2)$ charge is 
\begin{equation}
    Q=\sum_m i(c_{m,1}^\dagger c_{m,2}-c_{m,2}^\dagger c_{m,1}),
\end{equation}
where $(c_{m,1}, c_{m,2})$ transforms as a $O(2)$ vector. The most suitable quantum numbers for ED are diagonal and look like $\sum_{m\alpha}\alpha_{m\alpha}c^\dagger_{m\alpha}c_{m\alpha}$, so we transform with respect to the flavor index
\begin{equation}
    \mathbf{\ct}_m=\mathbf{U} \mathbf{c}_m,\qquad 
    \begin{bmatrix} \ct_{m,0} \\ \ct_{m,1} \\ \ct_{m,2} \end{bmatrix} =  \begin{bmatrix} 1 \\ & \frac{1}{\sqrt{2}} & \frac{i}{\sqrt{2}} \\ & \frac{1}{\sqrt{2}} & \frac{-i}{\sqrt{2}} \end{bmatrix} \begin{bmatrix} c_{m,0} \\ c_{m,1} \\ c_{m,2}\end{bmatrix}.
\end{equation}
In this basis, the conserved $O(2)$ charge is diagonal
\begin{equation}
    Q=\sum_m(\ct_{m,1}^\dagger \ct_{m,1} - \ct_{m,2}^\dagger \ct_{m,2}).
\end{equation}
Various fields are now expressed in the new basis as 
\begin{align}
    V^2&\mapsto\Vt_1^2+\Vt_2^2,&Q&=0, \nonumber \\
    V_{[a}V_{b]}&\mapsto\Vt_1\Vt_1-\Vt_2\Vt_2,&Q&=0, \nonumber \\
    V_{(a}V_{b)}&\mapsto(\Vt_1^2,\Vt_2^2),&Q&=\pm 2.
\end{align}

For general $O(N)$ model, we perform a similar transformation for the flavours $(1,2)$, $(3,4)$, \dots, and $(2n-1,2n)$ where $n=\lfloor N/2\rfloor$.
\begin{align}
    \mathbf{U}&=\left\{\begin{aligned}
        &\operatorname{diag}(1,\mathbf{U}_2,\dots, \mathbf{U}_2)&N&\in 2\mathbb{Z}\\
        &\operatorname{diag}(1,\underbrace{\mathbf{U}_2,\dots, \mathbf{U}_2}_{n=\lfloor N/2\rfloor},1)&N&\in 2\mathbb{Z}+1\\
    \end{aligned}\right.\nonumber\\
    \mathbf{U}_2&=\frac{1}{\sqrt{2}}\begin{bmatrix} 1 & i \\ 1 & -i \end{bmatrix},\qquad \mathbf{\ct}_m=\mathbf{U} \mathbf{c}_m,
\end{align}
where $\operatorname{diag}$ denotes a block-diagonal matrix. The diagonal charges are 
\begin{equation}
    Q_i=\sum_m(\ct^\dagger_{m,2i-1}\ct_{m,2i-1}-\ct^\dagger_{m,2i}\ct_{m,2i})\quad (i=1,\dots,n).
\end{equation}

\section{Implementing Symmetries in the Exact diagonalisation\label{app:ChoosingRepresentativeStates}}

In exact diagonalisation, implementing symmetries can block-diagonalise the Hamiltonian, divide the Hilbert space into sectors with different quantum numbers, and thus reduce the computational cost. The manifest global symmetry in our fuzzy-sphere model is 
\begin{equation*}
    G_N=U(1)_\text{charge}\times SO(3)_\text{rotation}\times O(N)_\text{flavour}.
\end{equation*}
In practice, we implement some $U(1)$ diagonal quantum numbers and discrete off-diagonal quantum numbers. 
\begin{itemize}
    \item Diagonal quantum numbers. They come from the Cartan sub-group of $G_N$ and are the maximal set of commuting operators that we can impose simultaneously.
    \item Off-diagonal quantum numbers. In certain sectors of the diagonal quantum numbers, we can implement some off-diagonal quantum numbers that do not commute with the Cartan sub-group. They come from the Weyl sub-group of $G_N$. Imposing these quantum numbers lifts some of the remaining degeneracies inside each weight space.
\end{itemize}
The representation of a state under a Lie group can be measured through the quadratic Casimir
\begin{equation}
    \langle\Phi|C_2|\Phi\rangle,\qquad C_2=\sum_i T_iT_i,
\end{equation}
where $|\Phi\rangle$ is an eigenstate and $T_i$ are the list of generators. 

We now discuss each part of the symmetry group individually.

\subsection{$U(1)$ charge symmetry}

We implement the electric charge conservation
\begin{equation}
    N_e=\sum_{m\alpha} c_{m\alpha}^\dagger c_{m\alpha}=N_\text{orb}.
\end{equation}

\subsection{$SO(3)$ sphere rotation symmetry}

We implement its Cartan sub-group, \emph{i.~e.}~the angular momentum conservation in the $z$-direction
\begin{equation}
    L_z=\sum_{m\alpha}m c_{m\alpha}^\dagger c_{m\alpha}.
\end{equation}
A spin-$l$ multiplet branches into components with $L_z=-l,-l+1,\dots,l-1,l$. To obtain all the $SO(3)$ representations with $l\in\mathbb{Z}$, it suffices to consider the $L_z=0$ sector. In this sector, we can implement its Weyl sub-group, \emph{i.~e.}~a $\mathbb{Z}_2$ that acts as the $\pi$-rotation along the $y$-axis. 
\begin{equation}
    \mathcal{R}_y:\ c_{m\alpha}\mapsto (-1)^{m+s}c_{-m,\alpha}.
\end{equation}

The spin $l$ of an eigenstate can be measured through the total angular momentum 
\begin{align}
    \langle L^2\rangle&=l(l+1),\nonumber\\
    L^2&=L_+L_-+L_z^2-L_z,\\
    L_\pm&=\sum_{m\alpha}\sqrt{(s\mp m)(s\pm m+1)}c^\dagger_{m\pm1,\alpha}c_{m,\alpha}.\nonumber
\end{align}

\subsection{Flavour Symmetry of the $O(2)$ Model}

We consider the flavour symmetry 
\begin{equation*}
    O(2) \cong U(1) \rtimes \mathbb{Z}_2,
\end{equation*}
Its irreducible representations are labelled by $U(1)$ charge $Q\in\mathbb{Z}$ and an improper $\mathbb{Z}_2$ quantum number only for $Q=0$.

We implement the conservation of $U(1)$ charge 
\begin{equation}
    Q=\sum_m\ct^{\dagger}_{m,1}\ct_{m,1}-\ct^{\dagger}_{m,1}\ct_{m,2}.
\end{equation}
Note that we work in the basis after transformation throughout the section (\textit{cf.}~Appendix \ref{app:complexfieid_basis}). The improper $\mathbb{Z}_2$ exchanges the two flavours
\begin{equation}
    \mathcal{Z}:\ \ct_{m,1}\leftrightarrow\ct_{m,2}.
\end{equation}
It reverses the $U(1)$ charge $\mathcal{Z}:Q\mapsto -Q$; the charge-0 sector is further divided into $\mathbb{Z}_2$-even and odd sectors. We list below the quantum numbers for several operators of interest
\begin{align}
    S&:&Q&=0&\mathcal{Z}&=+\nonumber\\
    J^\mu&:&Q&=0&\mathcal{Z}&=-\nonumber\\
    \phi&:&Q&=\pm 1\nonumber\\
    t&:&Q&=\pm 2
\end{align}
To obtain all the $O(2)$ representations, we go through $Q=0,\mathcal{Z}=\pm$ and $Q=1,2,\dots$.

\subsection{Flavour Symmetry of the $O(3)$ Model}

The $O(3)$ irreducible representations are labelled by $SO(3)$ spin $s\in\mathbb{Z}$ and an improper $\mathbb{Z}_2$ quantum number for each $s$. 

The Cartan sub-group of $SO(3)$ is $U(1)$ with charge
\begin{equation}
    S_z=\sum_m\ct^{\dagger}_{m,1}\ct_{m,1}-\ct^{\dagger}_{m,1}\ct_{m,2}.
\end{equation}
A $SO(3)$ spin-$s$ multiplet branches into components with $S_z=-s,-s+1,\dots,s-1,s$. To obtain all the $SO(3)$ representations, it suffices to consider the $S_z=0$ sector.

We implement in addition the Weyl reflection $\mathcal{Z}$ and the improper-$\mathbb{Z}_2$ $\mathcal{X}$
\begin{align}
    \mathcal{Z}&:\ \ct^{\dagger}_{m,1}\leftrightarrow\ct^{\dagger}_{m,2},\quad\ct^{\dagger}_{m,3}\mapsto-\ct^{\dagger}_{m,3}\nonumber,\\
    \mathcal{X}&:\ \ct^{\dagger}_{m,1}\leftrightarrow\ct^{\dagger}_{m,2}.
\end{align}
The $SO(3)$ spin-even representations are $\mathcal{Z}$-even and vice versa.

The $SO(N)$ quadratic Casimir admits a general definition 
\begin{equation}
    C_2=\sum_{a<b}\left[\frac{i}{\sqrt{2}}\sum_m c_{m,a}^\dagger c_{m,b}-c_{m,b}^\dagger c_{m,a}\right]^2.
\end{equation}
with the normalisation that the Dynkin index $T(\square)=1$ and $C_2(\square)=(N-1)/2$ for the vector representation. For $SO(3)$, it can be intepreted as half the total $SO(3)$ spin $\frac{1}{2}S^2$.
\begin{align}
    \langle S^2\rangle&=s(s+1),\nonumber\\
    S^2&=S_+S_-+S_z^2-S_z,\\
    S_+&=\sqrt{2}\sum_{m}\ct_{m,1}^\dagger \ct_{m,0}+\ct_{m,0}^\dagger \ct_{m,2},\nonumber\\
    S_-&=S_+^\dagger.\nonumber
\end{align}

We list below the quantum numbers for several operators of interest
\begin{align}
    S&:&s&=0&\langle S^2\rangle&=0&\mathcal{X}&=+&\mathcal{Z}&=+\nonumber\\
    J^\mu&:&s&=1&\langle S^2\rangle&=2&\mathcal{X}&=-&\mathcal{Z}&=-\nonumber\\
    \phi&:&s&=1&\langle S^2\rangle&=2&\mathcal{X}&=+&\mathcal{Z}&=-\nonumber\\
    t&:&s&=2&\langle S^2\rangle&=6&\mathcal{X}&=+&\mathcal{Z}&=+.
\end{align}

\begin{figure*}[t]
    \centering
    \includegraphics[width=0.8\linewidth]{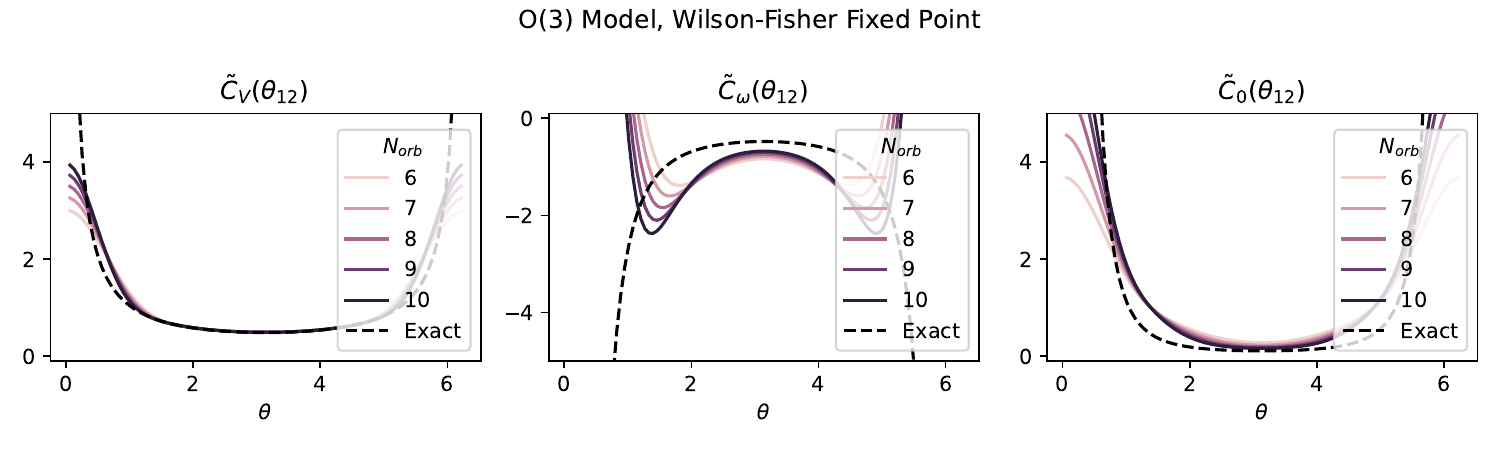}
    \includegraphics[width=0.8\linewidth]{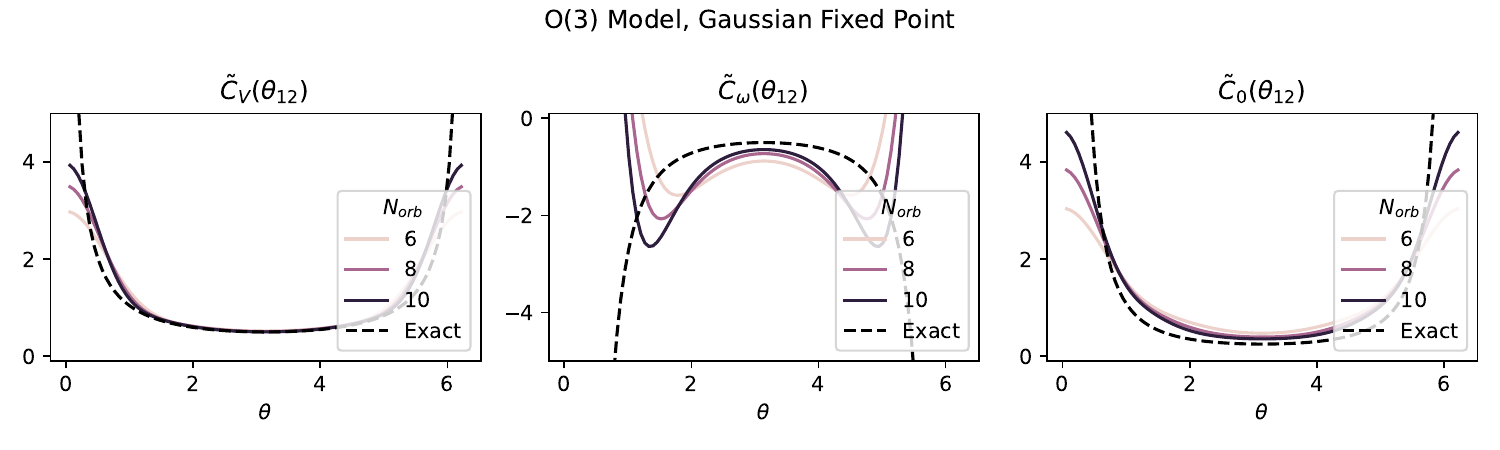}
    \caption{Two-point correlation functions of $\phi$, $\pi$, and $n_{00}$ operators in the $O(3)$ model at the WF and Gaussian fixed point. The CFT 2-point functions are dashed black line.  }
    \label{fig:corr_O3_WF}
\end{figure*}

\begin{figure*}[t]
    \centering
    \includegraphics[width=0.8\linewidth]{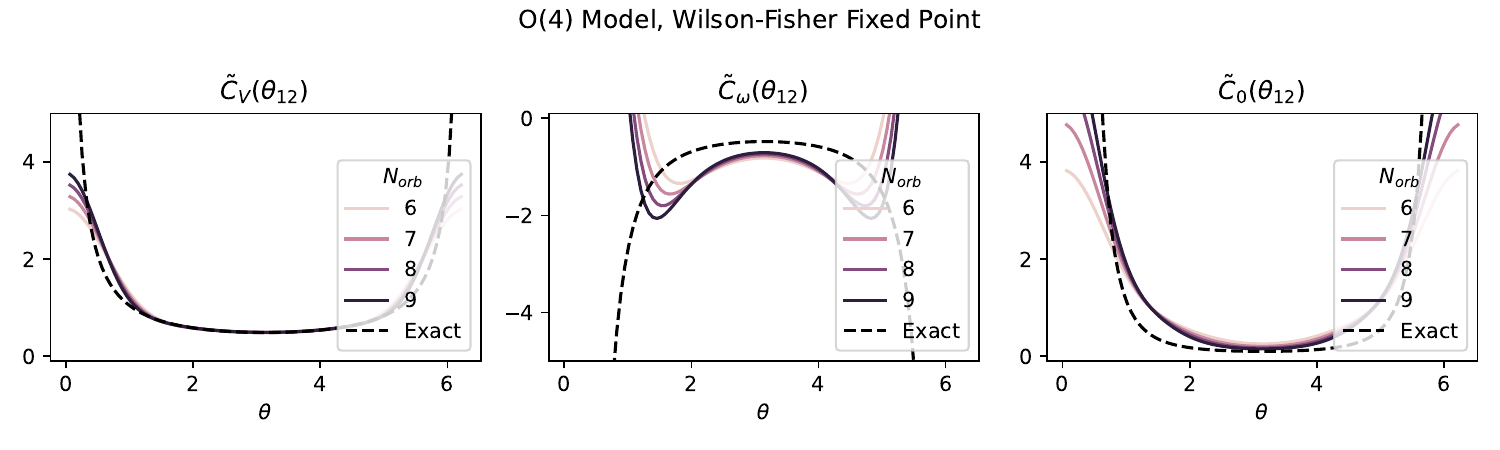}
    \includegraphics[width=0.8\linewidth]{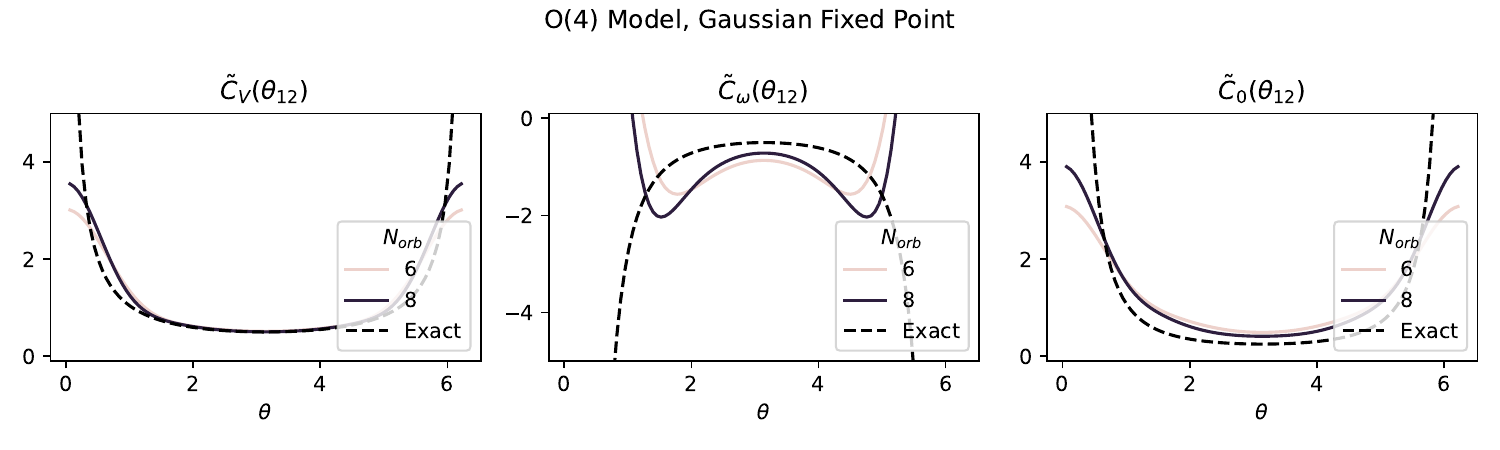}
    \caption{Two-point correlation functions of $\phi$, $\pi$, and $n_{00}$ operators in the $O(4)$ model at the WF and Gaussian fixed point. The CFT 2-point functions are dashed black line.  }
    \label{fig:corr_O4_WF}
\end{figure*}

\subsection{Flavour Symmetry of $O(4)$ model}

The connected part $SO(4)$ admits the well-known isomorphism:
\begin{equation}
    SO(4)\cong(SU(2)_L\times SU(2)_R)/\mathbb{Z}_2.
\end{equation}
Its irreducible representations are tensor-product representations of the two $SU(2)$'s, labeled by pairs of spins $(s_L,s_R)$ with $s_L,s_R \in \mathbb{Z}/2$. The projection of $\mathbb{Z}_2$ imposes an additional restriction $s_L+s_R\in\mathbb{Z}$. The improper-$\mathbb{Z}_2$ exchanges the two $SU(2)$'s. For $s_L\neq s_R$, it combines the two representations $(s_L,s_R)$ and $(s_R,s_L)$; for $s_L=s_R$, it divides the $(s_L,s_L)$ into the $\mathbb{Z}_2$-even and $\mathbb{Z}_2$-odd representations. 

The Cartan sub-group is the direct product $U(1)_L\times U(1)_R$ of the Cartan sub-groups of the two $SU(2)$'s. Its charges are 
\begin{align}
    S_{z,L}&=\frac{1}{2}\sum_m\ct^{\dagger}_{m,1}\ct_{m,1}-\ct^{\dagger}_{m,2}\ct_{m,2}+\ct^{\dagger}_{m,3}\ct_{m,3}-\ct^{\dagger}_{m,4}\ct_{m,4,}\nonumber\\
    S_{z,R}&=\frac{1}{2}\sum_m\ct^{\dagger}_{m,1}\ct_{m,1}-\ct^{\dagger}_{m,2}\ct_{m,2}-\ct^{\dagger}_{m,3}\ct_{m,3}+\ct^{\dagger}_{m,4}\ct_{m,4}.
\end{align}
An $SO(4)$ spin-$(s_L,s_R)$ multiplet branches into components with $S_{z,L}=-s_L,\dots,s_L$ and $S_{z,R}=-s_R,\dots,s_R$ respectively. To obtain all the $SO(4)$ representations, it suffices to consider the $(S_{z,L},S_{z,R})=(0,0)$ and $(\frac{1}{2},\frac{1}{2})$ sectors. 

\newcommand{\sm}{\hphantom{-}}

The Weyl sub-group of $O(4)$ is generated by 
\begin{align}
    \mathcal{Z}_1&:~\ct_{m,1}\leftrightarrow\ct_{m,2}\nonumber\\
    \mathcal{Z}_2&:~\ct_{m,3}\leftrightarrow\ct_{m,4}\nonumber\\
    \mathcal{X}&:~\ct_{m,1}\leftrightarrow\ct_{m,3},\quad \ct_{m,2}\leftrightarrow\ct_{m,4}.
\end{align}
They acts on the two quantum numbers as 
\begin{align}
    \mathcal{Z}_1&:&S_{z,L}&\mapsto-S_{z,R}&S_{z,R}&\mapsto-S_{z,L},\nonumber\\
    \mathcal{Z}_2&:&S_{z,L}&\mapsto\sm S_{z,R}&S_{z,R}&\mapsto \sm S_{z,L},\nonumber\\
    \mathcal{X}&:&S_{z,L}&\mapsto\sm S_{z,L}&S_{z,R}&\mapsto-S_{z,R}.
\end{align}
It is isomorphic to the dihedral group of a square $D_8$, one can see this by noting 
\begin{equation*}
    (\mathcal{Z}_1\mathcal{X})^4=1,\quad \mathcal{Z}_2(\mathcal{Z}_1\mathcal{X})=(\mathcal{Z}_1\mathcal{X})^{-1}\mathcal{Z}_2.
\end{equation*}
The Weyl sub-group of $SO(4)$ is $\mathbb{Z}_2\times\mathbb{Z}_2$ generated by $\mathcal{X}$ and $\mathcal{Z}_1\mathcal{Z}_2$. The $\mathcal{Z}_1$ or $\mathcal{Z}_2$ separately acts as the improper $\mathbb{Z}_2$. To obtain all the $O(4)$ representations, it suffices to consider the sectors 
\begin{align}
    (S_{z,L},S_{z,R},\mathcal{Z}_1,\mathcal{Z}_2,\mathcal{X})&=(0,0,+,+,\pm),&&(0,0,-,-,\pm)\nonumber\\
    &\hphantom{{}={}}(0,0,+,-,/),&&(\tfrac{1}{2},\tfrac{1}{2},/,\pm,/),
\end{align}
where $/$ means the quantum number is not conserved for the sector. The five choices of $\mathcal{Z}_1,\mathcal{Z}_2,\mathcal{X}$ for $(S_{z,L},S_{z,R})=(0,0)$ singles out respectively the $A_1,A_2,B_1,B_2$ and $E$ representation of $D_8$.

The Casimir is the sum of the total spins of the two $SU(2)$'s 
\begin{equation}
    \langle C_2\rangle=s_L(s_L+1)+s_R(s_R+1).
\end{equation}

We list below the quantum numbers for several operators of interest 
\begin{align}
    S&:&(s_L,s_R)&=(0,0)&\langle C_2\rangle&=0\nonumber\\
    &&(S_{z,L},S_{z,R})&=(0,0)\nonumber\\
    &&(\mathcal{X}_1,\mathcal{X}_2,\mathcal{Z})&=(+,+,+)\nonumber\\
    \phi&:&(s_L,s_R)&=(\tfrac{1}{2},\tfrac{1}{2})&\langle C_2\rangle&=\tfrac{3}{2}\nonumber\\
    &&(S_{z,L},S_{z,R})&=(\tfrac{1}{2},\tfrac{1}{2})\nonumber\\
    &&(\mathcal{X}_1,\mathcal{X}_2,\mathcal{Z})&=(/,+,/)\nonumber\\
    J^\mu&:&(s_L,s_R)&=(1,0),(0,1)&\langle C_2\rangle&=2\nonumber\\
    &&(S_{z,L},S_{z,R})&=(0,0)\nonumber\\
    &&(\mathcal{X}_1,\mathcal{X}_2,\mathcal{Z})&=(+,-,/)\nonumber\\
    t&:&(s_L,s_R)&=(1,1)&\langle C_2\rangle&=4\nonumber\\
    &&(S_{z,L},S_{z,R})&=(0,0)\nonumber\\
    &&(\mathcal{X}_1,\mathcal{X}_2,\mathcal{Z})&=(+,+,-)
\end{align}

\section{Suppelementary Results for Correlators}

We plot the dimensionless correlators~\eqref{eq:UV_corr1}--\eqref{eq:UV_corr3} for different system sizes for $O(N)$ ($N=3,4$) WF and free-scalar CFTs and compare them with the CFT results in Figs.~\ref{fig:corr_O3_WF} and \ref{fig:corr_O4_WF}.

We also append a derivation for Eq.~\eqref{eq:IR_corr2}. The momentum, transformed to the cylinder through the Weyl transformation $\bx=e^{\tau/R}\hat{\mathbf{n}}$, is 
\begin{equation}
    \pi(\br,\tau)_\text{cyl.}=i\left(\frac{r}{R}\right)^{\Delta_\phi+1}\left(\partial_r+\frac{\Delta_\phi}{r}\right)\phi,
\end{equation}
where the covariant derivative $D_\mu=\partial_\mu+\Delta B_\mu$ takes into account a modification to the connexion one-form~\cite{Weyl1921}
\begin{equation*}
    B=B_\mu\rd x^\mu=\rd\omega =(1/r)\,\rd r.
\end{equation*}
Now we consider the correlators 
\begin{align}
    \langle\phi(\bx_1)\phi(\bx_2)\rangle&=\frac{1}{(r_1^2+r_2^2-2r_1r_2\cos\theta_{12})^{\Delta\phi}}\\
    \langle\pi(\br_1)\pi(\br_2)\rangle&=\left.\frac{-1}{R^{2(\Delta_\phi+1)}}\left\langle (D_r\phi)(\bx_1)(D_r\phi)(\bx_2)\right\rangle\right|_{\substack{r_1=1\\r_2=1}}\nonumber\\
    &=\frac{-1}{R^{2(\Delta_\phi+1)}}\frac{2\Delta_\phi}{(2-2\cos\theta_{12})^{\Delta_\phi+1}}
    \label{eq:result1}
\end{align}
For the dimensionless correlator, a normalisation is needed
\begin{equation}
    \langle 0|\pi(\br)|\phi\rangle=\frac{i}{R^{\Delta_\phi+1}}\left.\left(\frac{\partial}{\partial r}+\frac{\Delta_\phi}{r} \right)\frac{1}{r^{2\Delta_\phi}}\right|_{r=1}=\frac{-i\Delta_\phi}{R^{\Delta_\phi+1}}.
\end{equation}
Hence, 
\begin{equation}
    \tilde{C}_\pi(\theta_{12})=\frac{\langle\pi(\br_1)\pi(\br_2)\rangle}{|\langle 0|\pi(\br)|\phi\rangle|^2}=\frac{-2/\Delta_\phi}{(2-2\cos\theta_{12})^{\Delta_\phi+1}}.
\end{equation}

%


\end{document}